\everydisplay{
	\abovedisplayskip=.3\baselineskip  plus.2ex minus.2ex\abovedisplayshortskip=-0.3\baselineskip plus.4ex minus.4ex
	\belowdisplayskip=.3\baselineskip plus.2ex minus.2ex\belowdisplayshortskip=.3\baselineskip plus.2ex minus.2ex
}
\documentclass[journal,9pt]{IEEEtran}
\ifCLASSINFOpdf
\else
\fi

\usepackage[T1]{fontenc}
\usepackage[latin9]{inputenc}
\usepackage{array}
\usepackage{float}
\usepackage{units}
\usepackage{multirow}
\usepackage{amsmath}
\usepackage{amssymb}
\usepackage{graphicx}
\usepackage{subfigure}
\usepackage{esint}
\usepackage{xcolor}
\usepackage{epstopdf}
\usepackage{arydshln}
\usepackage{cases}
\allowdisplaybreaks[4]
\usepackage{float}
\usepackage[square, comma, sort&compress, numbers]{natbib}  
\makeatletter

\floatstyle{ruled}
\newfloat{algorithm}{tbp}{loa}
\providecommand{\algorithmname}{Algorithm}
\floatname{algorithm}{\protect\algorithmname}

\makeatother

\usepackage[english]{babel}

\makeatletter
\adddialect\l@ENGLISH\l@english
\makeatother

\begin{document}
	\linespread{0.9}
%
\title{\hspace*{-2.8mm} \LARGE \bf Resilient Global Practical Fixed-Time Cooperative Output Regulation of Uncertain Nonlinear Multi-Agent Systems 
	Subject to Denial-of-Service Attacks}
%
%

%

\author{Wenji~Cao, Lu~Liu,~\IEEEmembership{Senior Member,~IEEE,} Zehua~Ye, Dan~Zhang,~\IEEEmembership{Senior Member,~IEEE,} Gang~Feng,~\IEEEmembership{Fellow,~IEEE}
	\thanks{Wenji Cao, Lu Liu and Gang Feng are with the Department of Biomedical Engineering, City University of Hong Kong, Kowloon, Hong Kong, China (e-mail: wenjicao2-c@my.cityu.edu.hk(W.Cao); luliu45@cityu.edu.hk(L.Liu); megfeng@cityu.edu.hk(G.Feng))}
\thanks{Zehua Ye is with the Research Center of Automation and Artificial Intelligence, Zhejiang University of Technology, Hangzhou, China (e-mail: 2111803109@zjut.edu.cn(Z.Ye))}
\thanks{Dan Zhang is with the Research Center of Automation and Artificial Intelligence, and State Key Laboratory of Green Chemical Synthesis and Conversion, Zhejiang University of Technology, Hangzhou, China (danzhang@zjut.edu.cn(D.Zhang))}
	\thanks{This work was supported by the Research Grants Council of Hong Kong under Projects CityU/11205024 and CityU/11207323, the National Nature Science Foundation of China under Grants 62222318/62322315, the Fundament Research  for the Provincial Universities of Zhejiang under Grant RF-C2024002, and the Zhejiang Provincial Natural Science Foundation of China under Grant LR22F030003.
	(\textit{Corresponding author: Gang~Feng})}
}
\maketitle


\begin{abstract}
This paper investigates the {problem of resilient global practical fixed-time cooperative output regulation of} uncertain nonlinear multi-agent systems subject to denial-of-service attacks.
A novel distributed resilient adaptive fixed-time control strategy is proposed, which consists of a novel distributed resilient fixed-time observer {with} a chain of nonlinear filters and a novel distributed resilient adaptive fixed-time controller.
It {is} shown that the {problem of resilient global practical} fixed-time cooperative output regulation can be solved by the proposed control strategy.
More specifically, the proposed {distributed} control strategy ensures the global boundedness of all the signals in the resulting closed-loop system and the global convergence of the regulated outputs to a {tunable} residual set in a fixed time.
A simulation example is finally provided to illustrate the efficacy of the proposed control strategy.
\end{abstract}


\begin{IEEEkeywords}
	Cooperative output regulation, nonlinear multi-agent systems, fixed-time control, denial-of-service attacks.     
\end{IEEEkeywords}

%
\IEEEpeerreviewmaketitle

\newtheorem{lemma}{Lemma}
\newtheorem{theorem}{Theorem}
\newtheorem{assumption}{Assumption}
\newtheorem{remark}{Remark}
\newtheorem{corollary}{Corollary}
\newtheorem{definition}{Definition}
\newtheorem{problem}{Problem}
\newtheorem{property}{Property}

\section{Introduction}\label{sec:1}
Cooperative control of multi-agent systems (MASs) has attracted {considerable research interest} in the past few decades due to their {wide} applications in the real world ranging from formation control of mobile robots to optimal deployment of sensor networks \citep{jadbabaie2003coordination,anderson2008uav,ren2008distributed,ren2011distributed}. 
The topics on cooperative control of MASs include but not limited {to consensus}, cooperative output regulation (COR), formation, and containment.
In particular, the COR problem of MASs is widely studied and numerous remarkable results have been obtained \citep{su2011cooperative,li2016distributed,xu2014distributed,dong2018cooperative}.
For example, the COR problem of MASs with linear {agent} dynamics is addressed by {distributed feedforward methods} \citep{su2011cooperative,li2016distributed}. 
In \cite{xu2014distributed} and \cite{dong2018cooperative}, the same problem is addressed for nonlinear MASs in strict-feedback normal form and for second-order nonlinear MASs, respectively. 
However, in the aforementioned works on COR of MASs, there is a common assumption that communication networks of MASs are free of any cyber-attacks.

In practical applications, communication networks of MASs often suffer from cyber-attacks, which can cause {their performance degradation and even operation failures.}
Typically, cyber-attacks can be classified into two types: denial-of-service (DoS) attacks \citep{de2015input} and deception attacks \citep{pasqualetti2013attack}. 
In recent years, {cooperative control of MASs subject to DoS attacks has attracted increasing} interest and abundant results on this topic have been reported \cite{feng2016distributed,hu1,lu2018distributed,feng2019secure,deng2020mas,deng2022resilient,zhangdananddengchao,li2023111250}. 
For example, the consensus problem of linear MASs under DoS attaks is investigated in \citep{lu2018distributed,feng2019secure}. 
{In \cite{deng2020mas,deng2022resilient}, the authors address the problem of COR for linear MASs {under} DoS attacks via the distributed resilient control schemes.}
{The} resilient COR problem is also investigated for nonlinear MASs {under} DoS {attacks \cite{zhangdananddengchao}}.
It is noted that the aforementioned works on cooperative control of MASs {under} DoS attacks focus on either asymptotical or exponential convergence. 
In other words, the coordination tasks are accomplished only when time goes to infinity.

However, in practice, many engineering applications in cooperative control of MASs call for finite-time or fixed-time convergence instead of exponential or asymptotical convergence.
In recent years, researchers have paid much attention to solve finite-time cooperative control problems of MASs \citep{wang2010finite,li2011finite,zuo2014new,li2018finite,li2022event,sarrafan2022resilient,cao2023cooperative,ZHANG2025112062}. 
In particular, several finite-time consensus control {schemes} have been designed for MASs {with integrator type agent dynamics \cite{wang2010finite,li2011finite,zuo2014new} and with strict-feedback nonlinear agent} dynamics \cite{li2022event, li2018finite}.
Most recently, {finite-time cooperative control problems} of MASs under DoS attacks {are} studied in \cite{sarrafan2022resilient,cao2023cooperative,ZHANG2025112062}.
{However, those finite-time cooperative control schemes suffer from one undesirable disadvantage, that is, the upper bounds of their settling times depend on their initial conditions, which might tend to be very large if the initial conditions deviate far away from the equilibrium points.
Inspired} by fixed-time stability theory for single systems \citep{polyakov2011nonlinear}, fixed-time cooperative control of MASs has been {studied} \citep{tian2018fixed,zuo2017fixed, xu2021novel,zhan2021adaptive,zhang2019bipartite, du2020distributed,you2020fixed,10643866,zhang2022cooperative,10453626,zhoushiyu}.
In particular, fixed-time consensus {problems} of MASs {have been investigated for integrator type agent dynamics \citep{tian2018fixed,zuo2017fixed,xu2021novel}, linear agent dynamics \citep{zhang2019bipartite,zhan2021adaptive}, and nonlinear agent} dynamics \citep{du2020distributed,you2020fixed,10643866}.
{The problem of fixed-time COR for linear MASs is solved by a distributed event-triggered control strategy in \cite{zhang2022cooperative}, while the same problem for nonlinear MASs is solved by  a
smooth distributed adaptive controller in \cite{10453626}.}
{{More recently}, a few works study the fixed-time consensus {problem} of MASs under DoS attacks \cite{yang2020observer,xu2024fixed}. {However,} the consensus can only be achieved {in \cite{yang2020observer}} for those agents that have access to the leader's information for the concerned high-order MAS under connectivity-broken attacks.
In \cite{xu2024fixed}, the resilient fixed-time consensus {problem of a special class of} nonlinear MASs is {addressed via a distributed fuzzy control strategy, where the whole communication network is  assumed to be} blocked during the attack time intervals. 
In fact, to our best knowledge, there is no reported result in open literature on fixed-time COR of MASs under DoS attacks, which motivates this study.}

This paper {investigates the problem of resilient global practical fixed-time COR} for uncertain nonlinear MASs subject to DoS attacks. 
{It is {well known} that the adaptive backstepping control technique offers a systematic framework for designing controllers for uncertain nonlinear systems in parametric strict-feedback form, {though it requires} {the higher order derivatives of the reference signal to be} continuous \citep{krstic1995nonlinear}. However, the derivatives of the reference signals generated by {most} existing distributed fixed-time observers are discontinuous due to the disruption in communication among agents caused by DoS attacks. 
Thus, the main challenge in solving the problem is how to design the distributed fixed-time control protocol for {uncertain} nonlinear MASs under DoS attacks to ensure the {fixed-time stability of the resulting nonlinear} closed-loop system and the global fixed-time convergence of the regulated outputs.}

{To address the above-mentioned challenge, we propose a novel distributed resilient adaptive fixed-time control strategy. {In particular, a} novel distributed resilient fixed-time observer {with} a chain of nonlinear filters is developed for each agent to estimate the state of the exosystem, whose second order to {$n$th} order derivatives are {thus} guaranteed to {be continuous} even under DoS attacks. Our idea is inspired by {\cite{zhangdananddengchao,li2023111250}} where a chain of the first-order linear filters are introduced to achieve the {similar} objective. However, our work differs from that of {\cite{zhangdananddengchao,li2023111250}} in two aspects. That is, our filters are nonlinear and the estimation errors converge to zero within a fixed time while the filters in {\cite{zhangdananddengchao,li2023111250}} are linear and the estimation errors converge {to zero} {only} exponentially. It is shown that the state of the observer globally approaches to the state of the exosystem under some mild conditions in a fixed time, whose upper bound is independent of {initial conditions} and is explicitly given}. 
Then, based on the adaptive backstepping technique and the proposed observer, a novel distributed resilient adaptive fixed-time control strategy is developed. It {is} shown that the proposed control strategy ensures the global boundedness of all the signals in {the resulting} closed-loop system and the global convergence of the regulated outputs to a {tunable} residual set in a fixed time{, whose upper bound is also independent of  initial conditions and is explicitly} given.
To our {best} knowledge, such a {problem of} resilient global practical fixed-time COR for uncertain nonlinear MASs subject to DoS attacks is addressed {for} the first time.

\emph{Notation}: $\mathbb N_+$, $\mathbb R_+$ and $\mathbb R$ denote the sets of positive natural numbers, positive real numbers, and real numbers, respectively.
Let $\mathbb R^n$ and $\mathbb R^{n\times n}$ be the sets {of $n$--dimensional real vectors and $n \times n$--dimensional} matrices, respectively.
 $\|\cdot\|$ denotes the Euclidean norm for vectors.
{For a vector $x\!=\![x_1,x_2,\cdots,$ $x_n]^T$ and a constant $c\geq 0$, denote $x^c=[x_1^c,x_2^c,\cdots,x_n^c]^T$.}
$I_n\!\in\! \mathbb R^{n\times n}$ denotes the identify matrix. 
Denote $\otimes$ as the Kronecker operator for matrices. 
Let $A^T$ be the transpose of matrix $A${, and $A$ is said to be} symmetric if $A^T=A$.
{$A >0$ $(A\geq 0)$ represents that $A$ is positive definite ({$A$} is positive semi-definite).}
Let $\lambda_{min}(A)$ and $\lambda_{max}(A)$ be the smallest and largest eigenvalues of a symmetric matrix $A$, {respectively.}
sign$(\cdot)$ denotes a sign function.  
Given two sets $\Upsilon$ and $\Phi$, the relative complement of $\Phi$ in $\Upsilon$ is denoted as $\Upsilon\backslash \Phi$.
\section{Preliminaries and Problem Formulation}
\subsection{Graph Theory}
Consider a MAS with $N$ agents {and an exosystem}. The communication network among $N$ agents {is} modeled by an undirected graph 
$\mathcal{G}=(\mathcal V,\mathcal{E})$, {where} $\mathcal V=\{1,2,\cdots,N\}$ represents the node set and $\mathcal E=\{(j,i):i,j\in \mathcal V\}$ represents the edge set. 
A path from node $i$ to node $j$ is a sequence of ordered edges $(i, {j_1}),({j_1}, {j_2}),\cdots,({j_k}, j)$. 
$\mathcal G$ is called a connected graph, if there exists a path between every pair of nodes. Define the adjacency matrix associated with graph $\mathcal{G}$ as follows: $\mathcal A=[a_{ij}]\in \mathbb{R}^{N\times N}$, where $a_{ii}=0$, $a_{ij}=1$ if $(j,i)\in \mathcal E$ and $a_{ij}=0$, otherwise.
Define the elements $l_{ij}$ of {the} Laplacian matrix $\mathcal L\in \mathbb{R}^{N \times N}$ as follows: $ l_{ii}=\sum_{j={1}}^N a_{ij}$ and $ l_{ij}=-a_{ij}$ if $i\neq j$.
Denote node $0$ as the exosystem.
Then, we introduce graph $\mathcal {\bar G}=(\bar {\mathcal V},\bar {\mathcal{E}})$ to model the interaction among $N$ agents and {the} exosystem, where $\bar {\mathcal V}=\{0,1,\cdots,N\}$ and $\bar {\mathcal E}=\{(j,i):i,j\in \bar{\mathcal V}\}$. Node $0$ is said to be globally reachable if node $0$ has directed paths in $\bar{\mathcal G}$ to all the nodes in $\mathcal G$.  
Let $a_{i0}$ be the weight of {a directed edge} from the
exosystem to agent $i$, $i=1,2,\cdots, N$.
$a_{i0}=1$, {if} agent $i$ has access to information from the exosystem; $a_{i0}=0$, otherwise. 
Denote $\mathcal H = diag(a_{10}, a_{20}, \cdots,  a_{N0})+\mathcal L$.

\subsection{DoS Attacks}
To denote whether edge $(j,i)$ {in graph $\bar{\mathcal G}$} is under {attack} or not, variables $\sigma_{ij}(t)$, $(j,i)\in \bar {\mathcal E}$ are introduced.
If edge $(j,i)$ is under {attack}, let $\sigma_{ij}(t)=0$, and otherwise $\sigma_{ij}(t)=1$.
Then, we introduce a time-varying graph $\mathcal {\bar G}^{\sigma(t)}$ to model the {communication} topology among $N$ agents and {the} exosystem in consideration of DoS attacks.
Define the elements $a_{ij}^{\sigma(t)}$ of the adjacency matrix $\mathcal {\bar A}^{\sigma(t)}\in \mathbb R^{(N+1)\times(N+1)}$ associated with {graph} $\mathcal {\bar G}^{\sigma(t)}$ as follows: $a_{ij}^{\sigma(t)}\!=\!\sigma_{ij}(t)\cdot a_{ij}$, for $(j,i)\in \bar {\mathcal E}$.
Graph $\mathcal {\bar G}^{\sigma(t)}$ is said to be under DoS attacks if at least one of its edges is attacked.
Let $t_0=0$ be the initial time. Denote $[t_k^a, t_k)$, $k\in  \mathbb N_+$ as the time {interval of the {$k$th} DoS attack} for graph $\mathcal {\bar G}^{\sigma(t)}$, {where $t_k^a$ is the starting instant and $t_{k}$ is the ending instant.}
Denote the set of time intervals for graph $\mathcal {\bar G}^{\sigma(t)}$ under DoS attacks as 
\small$$\Psi_D(t_0,t)\!=\!\left\{\!\!
\begin{array}{l}
	\bigcup_{l\!=\!1}^{k\!-\!1}[t_l^a,t_l),t\!\in\! [t_{k-1}, t^a_k),\\
	\bigcup_{l\!=\!1}^{k\!-\!1}[t_l^a,t_l)\bigcup[t_k^a,t],t\!\in\! [t_k^a,t_{k}),
\end{array}\right.$$\normalsize
{and} its length is {denoted as $|\Psi_D(t_0,t)|$}.
Let the set of time intervals for graph $\mathcal {\bar G}^{\sigma(t)}$ without DoS attacks be $\Psi_N(t_0,t)=[t_0,t]\backslash \Psi_D(t_0,t),$ and its length is {denoted as
$|\Psi_N(t_0,t)|$.}

\subsection{Useful Definitions and Lemmas}
Consider a system given below,
\begin{equation}\label{finiteq}
	\dot x(t)=f(t,x), x(0)=x_0,
\end{equation}
where $x(t)\!\in\! \mathbb R^n$ is its state, $f\!:\mathbb R_+\!\times \mathbb R^n\!\to\! \mathbb R^n$ denotes a nonlinear function, which can be discontinuous. The solutions of {system} (\ref{finiteq}) are understood in the sense of Filippov \citep{filippov2013differential}. Assume that $x(t)=0$ is an equilibrium {point} of {system} (\ref{finiteq}).
Denote $\bar x(t,x_0)$ as the solution of  {system} (\ref{finiteq}).  
The equilibrium {point} $x(t)=0$ of {system} (\ref{finiteq}) is said to be globally finite-time convergent if for any $x_0\in\mathbb R^n$,
there is a function $T(x_0)$ such that $\lim\limits_{t\to T(x_0)}\!\bar x(t,x_0)\!=\!0$.
\begin{definition}	\cite{bhat2000finite,polyakov2011nonlinear,zhu2011attitude} \label{stable} 
The equilibrium point $x(t)=0$ of {system} (\ref{finiteq}) is said to be
	
	(i) globally finite-time stable if it is globally asymptotically stable and globally finite-time convergent;
	
	(ii) globally fixed-time stable if it is globally finite-time stable and there is $T\!>\!0$ such that $T(x_0)\!<\!T$, for any $\!x_0\!\in \mathbb R^n$;
	
	(iii) globally practically fixed-time stable if there exist $r\!>\!0$ and $T\!>\!0$ such that $\|\bar x(t,x_0)\|\!\leq\! r$, for all $t\!\geq\! T$ {and any $x_0\!\in\! \mathbb R^n$}.
	\end{definition}

\begin{lemma}\label{fixed time stability}\citep{polyakov2011nonlinear}
	If there exists a continuous radially unbounded and positive definite function $V(x)$ satisfying $
		\dot V(x)\!\leq\! -c_1(V(x))^p\!-c_2(V(x))^q,$ where {$c_1, c_2\!>\!0$}, $p\!\in\! (0,1)$ and $q\!\in\!(1,\infty)${, then} the equilibrium point $x(t)\!=\!0$ of {system} (\ref{finiteq}) is globally fixed-time stable and the upper bound for its convergence time is given by ${T(x_0)}\!\leq\! \frac{1}{c_1(1-p)}\!+\!\frac{1}{c_2(q-1)}$. 
\end{lemma}
\begin{lemma}\citep{deng1997stochastic}\label{lemma1}
	(Young's inequality) For $p, q\in \mathbb R$, inequality $pq\!\leq\! \frac{l^a}{a}|p|^a\!+\!\frac{1}{bl^b}|q|^b$ holds,
{where $a,b\!>\!1$, $l>0$ and $(b-1)(a-1)=1$.}
\end{lemma}

\begin{lemma}\label{lem1}\citep{ni2010leader}
	The matrix $\mathcal H$ associated {with} the graph $\mathcal{\bar G}$ has the following properties: (i) {$\mathcal H\!\geq\! 0$}; {and} (ii) {$\mathcal H\!>\!0$} if and only if node $0$ is globally reachable {in graph} $\bar{\mathcal G}$.
\end{lemma}
\begin{lemma}\label{le1}\citep{zuo2014new}
{Given} $x_i\!\geq\!0$, one has $\sum_{i\!=\!1}^{N}\!x_i^p\!\geq\! (\sum_{i\!=\!1}^{N}\!x_i)^p$ if $p\in(0,1]$,
	and
	$\sum_{i=1}^{N}x_i^q\geq N^{1-q}(\sum_{i=1}^{N}x_i)^q$ if $q\in(1,\infty)$.
\end{lemma}
\begin{lemma}\label{le2}\citep{song2021distributed} 
{Given} $x_{ij}\geq0$, one has $\sum_{i=1}^{N}\sum_{j=1}^{N}\!x_{ij}^p\geq (\sum_{i=1}^{N}\sum_{j=1}^{N}x_{ij})^p$ if $p\in(0,1]$, 
	and
	$\sum_{i=1}^{N}\sum_{j=1}^{N}x_{ij}^q\geq N^{2(1-q)}(\sum_{i=1}^{N}\sum_{j=1}^{N}x_{ij})^q$ if $q\in(1,\infty)$.
\end{lemma}
\subsection{Problem Formulation}
	Consider {an uncertain nonlinear MAS with $N$ agents.} {For agent $i$, its dynamics are described in the following lower triangular parametric strict-feedback form \citep{krstic1995nonlinear},}
\begin{small}\begin{subequations}\label{agent}
\begin{align}
	\dot x_{is}&=x_{i(s+1)}+\phi_{is}^T(t,\bar x_{is})\theta_i+b_{is}v, s\!=\!1,\!\cdots\!,n\!\!-\!\!1,\\
	\dot x_{in}&=u_i+\phi_{in}^T(t,x_i)\theta_i+b_{in}v,\\
	e_i&=x_{i1}+R_iv, {i=1,2,\cdots,N,}
\end{align}
\end{subequations}\end{small}where $\bar x_{is}\!=\![x_{i1}, x_{i2},\cdots, x_{is}]^T\! \in\! \mathbb R^s$, $s\!=\!1,\!\cdots\!,n-1$, $x_i=[x_{i1},$ $ x_{i2},\cdots, x_{in}]^T \!\in\! \mathbb R^n$ is its state vector, $u_i\!\in\! \mathbb R$ and $e_i\!\in\! \mathbb R$ are its input and regulated output, respectively{;} $\theta_i\in \mathbb R^m$ denotes an {unknown} constant vector{,}
$\phi_{is}(t,\bar x_{is})\in \mathbb R^m$ is a known smooth nonlinear function{,}
$b_{is}^T, R_i^T\!\in\! \mathbb R^q$ are known constant vectors{;} and $v\!\in\! \mathbb R^q$ represents the exogeneous signal 
generated by the exosystem,
\begin{small}\begin{equation}\label{exo}
	\dot v=Sv,
\end{equation}\end{small}with $S\in \mathbb R^{q\times q}$.

{To proceed, the following assumptions are introduced.}
\begin{assumption}\label{assdos}
	(DoS Duration) There exist positive constants $\nu_d $ and $p_d$ with $p_d >1$ such that 
	\begin{small}\begin{equation}
			|\Psi_D(t_0,t)| \leq \frac{t-t_0}{p_d} + \nu_d, \forall t  \geq t_0.
	\end{equation}\end{small}
\end{assumption}
\begin{assumption}\label{ass1}
	Node $0$ is globally reachable {in graph} $\bar{\mathcal G}$.
\end{assumption}
\begin{assumption}\label{assmatrix}
	The eigenvalues of $S$ are semi-simple with zero real parts.
\end{assumption}

\begin{remark}
{Assumption \ref{assdos} guarantees the boundedness of DoS duration,} which is also {adopted in} \cite{de2015input}.
	Assumption \ref{ass1} implies that there is at least one agent who has access to the state of the exosystem \cite{deng2022resilient,sarrafan2022resilient,cao2023cooperative}.
	Assumption \ref{assmatrix} is commonly adopted as in \cite{dong2018cooperative} and \cite{zhangdananddengchao}, which guarantees the boundedness of the exosystem state.
\end{remark}

The fixed-time COR problem for the uncertain nonlinear MAS (\ref{agent}) and exosystem (\ref{exo}) under DoS attacks investigated in this paper is formally stated as follows:

\begin{definition}(Resilient Global Practical Fixed-Time COR Problem)
	Given MAS (\ref{agent}) and exosystem (\ref{exo}) subject to DoS attacks, design a distributed resilient fixed-time control scheme such that for any initial condition,\\ {(a) all the signals in the resulting closed-loop system are globally bounded, {and}\\
	(b) the regulated output globally converges to a residual set in a fixed time, i.e., $e_{i}\in \Theta$, $\forall t\geq t_a$, where $t_a$ is a fixed time and $\Theta \triangleq \{e_i: |e_i|\leq \varepsilon\}$ is defined as a residual set with $\varepsilon$ being a positive constant.}
\end{definition}
\begin{remark}
{The size of the residual set $\Theta$ can be {tuned by} the design parameters, and the guidelines of tuning will be provided in the sequel.}
 	\end{remark}

\section{Main Results}
In this section, a distributed resilient fixed-time observer {with} a chain of nonlinear filters is designed for each agent to estimate the state of the exosystem by taking into account of DoS attacks. Then, based on the adaptive backstepping technique and the {proposed} observer, a novel distributed resilient fixed-time control strategy is developed to address the resilient global practical fixed-time COR problem described in the previous section. 
\subsection{Distributed Resilient Fixed-Time Observer}
{As the so-called adaptive backstepping technique relies on the reference signal and its up to $n${-}th derivative {being} continuous, the following distributed resilient fixed-time observer {with} a chain of nonlinear filters is designed to estimate the state of the exosystem for agent $i$,}
\begin{small}\begin{subequations}\label{ob}
	\begin{align}
		\dot {\hat v}_i=&S\hat v_i-\delta_1(\hat v_i-\epsilon_{i1})\!-\!\delta_2(\hat v_i-\epsilon_{i1})^{\frac{a}{b}}\!-\!\delta_3(\hat v_i-\epsilon_{i1})^{2-\frac{a}{b}},\label{obb1}\\
		\dot{\epsilon}_{ik}=&S\epsilon_{ik}-\delta_1(\epsilon_{ik}-\epsilon_{i(k+1)})\!-\!\delta_2(\epsilon_{ik}-\epsilon_{i(k+1)})^{\frac{a}{b}}\nonumber\\
		&-\delta_3(\epsilon_{ik}-\epsilon_{i(k+1)})^{2-\frac{a}{b}}, k=1,2,\cdots,n-1,\label{obb4}\\
		\dot\epsilon_{in}=&S\epsilon_{in}-\mu_1\!\sum_{j=0}^{N}\!a_{ij}^{\sigma(t)}\!(\epsilon_{in}-\epsilon_{jn})-\mu_2\!\sum_{j=0}^{N}\!a_{ij}^{\sigma(t)}\!(\epsilon_{in}-\epsilon_{jn})^{\frac{a}{b}}\nonumber\\[-2mm]
		&-\mu_3\sum_{j=0}^{N}a_{ij}^{\sigma(t)}\!(\epsilon_{in}-\epsilon_{jn})^{2-\frac{a}{b}},\label{obb2}
	\end{align}
\end{subequations}\end{small}where $\hat v_i$, $\epsilon_{ik}\in \mathbb R^q$, $k=1,2,\cdots,n,$ are the states of the observer, $\epsilon_{0n}=v$, $a$, $b$, $\delta_1$, $\delta_2$, $\delta_3$, $\mu_1$, $\mu_2$ and $\mu_3$ are positive constants.

Define elements $\bar l_{ij}, \tilde l_{ij}$ of {two} matrices $\mathcal {\bar L}, \mathcal {\tilde L} \in \mathbb{R}^{N \times N}$ as follows: $\bar l_{ii}=\sum_{j=1}^N \bar a_{ij}$ and $\bar l_{ij}=-\bar a_{ij}$ for $i\neq j$, and 
$\tilde l_{ii}=\sum_{j=1}^N \tilde a_{ij}$ and $\tilde l_{ij}=-\tilde a_{ij}$ for $i\neq j$, where $\bar a_{ij}=(a_{ij})^{\frac{2b}{a+b}}$ and $\tilde a_{ij}=(a_{ij})^{\frac{2b}{3b-a}}$.
Let $\mathcal {\bar H}=diag(\bar a_{10}, \bar a_{20}, \cdots, \bar a_{N0})+\mathcal {\bar L}$ and $\mathcal {\tilde H}=diag(\tilde a_{10}, \tilde a_{20},\cdots,  \tilde a_{N0})+\mathcal {\tilde L}$, where $\bar a_{i0}=(a_{i0})^{\frac{2b}{a+b}}$ and $\tilde a_{i0}=(a_{i0})^{\frac{2b}{3b-a}}$.

Now, {the following} key technical result on the  proposed distributed resilient fixed-time observer is presented. 
\begin{theorem}\label{lemmat}
 	 Suppose Assumptions {\ref{assdos}--\ref{ass1}} hold. For any initial {condition, the state $\hat v_i$} of the distributed resilient observer (\ref{ob}) globally converges to the state {$v$} of the exosystem (\ref{exo}) in a fixed time if there exist positive constants $\delta_1$, $\mu_1$, $\mu_2$, $\mu_3$, $c_s$, positive odd integers $a$, $b$ {with} $a<b$ {such that}

	1) {\small$\lambda_{\max}(S+S^T)-2\mu_1\lambda_{\min}(\mathcal H)<0$\normalsize,}
	
	2) \small$c_1(p_d-1)-c_4>0$\normalsize,
	
	3) {$c_4e^{\frac{c_4(b-a)}{2b}\nu_d}-c_2(p_d-1)< 0$}\normalsize,
	
	4) \small$e^{\frac{c_4(b-a)}{2b}(\frac{\tilde t_o-\bar t_o}{p_d}+\nu_d)}-\frac{c_2(b-a)}{2b}(\tilde t_o-\bar t_o-\frac{\tilde t_o-\bar t_o}{p_d}-\nu_d)\leq 0$\normalsize,
	
	5) \small$\lambda_{max}(S+S^T)\leq 2\delta_1$\normalsize,\\
	where {$c_1\!=\!2\mu_1\lambda_{\min}(\mathcal H)-\lambda_{\max}(S+S^T)$,} $c_2\!=\!\frac{\mu_2}{2}\!(4\lambda_{min}\!(\mathcal{\bar H}))\!^{\frac{a\!+\!b}{2b}}$,
$c_3\!=\!\frac{\mu_3}{2}\!(2qN^2)\!^{\frac{a\!-\!b}{2b}}(\!4\lambda_{min}\!(\mathcal {\tilde H}))\!^{\frac{3b\!-\!a}{2b}}$, 
$c_4\!=\!\max\{\lambda_{max}(S+S^T),c_s\}$, $\bar t_o$ and $\tilde t_o$ are calculated by 
	\begin{small}\begin{align}\label{to1}
{c_3}(e^{\frac{(b-a)(c_1(p_d-1)-c_4)}{2bp_d}t-\frac{(c_1+c_4)(b-a)}{2b}\nu_d}-1)-{c_1}=0,
	\end{align}\end{small}and
	\begin{small}\begin{align}\label{ton}
{c_4(b-a)}e^{\frac{c_4(b-a)}{2b}(\frac{t-\bar t_o}{p_d}+\nu_d)}-{c_2(b-a)(p_d-1)}=0,
	\end{align}\end{small}respectively.
Moreover, an upper bound of the settling time is given by 
	\begin{small}\begin{align}\label{to}
		t_{o}=\tilde t_o+{\frac{2bn}{b-a}(\frac{1}{\delta_2}+\frac{1}{\delta_3(Nq)^{\frac{a-b}{2b}}})}.
	\end{align}\end{small}
\end{theorem}
\textbf{\textit{Proof:}}
Please see Appendix.

The {distributed} observer designed in (\ref{ob}) guarantees that the state of the observer $\hat v_i$ and its derivatives $\hat v_i^{(1)}$, $\hat v_i^{(2)}$, $\cdots$, $\hat v_i^{(n)}$ are continuous. This property is {essential} in the design of the distributed resilient adaptive fixed-time control strategy {to} be given in next subsection.
\begin{remark}
{{It can be seen from conditions 1)--5) that the choices} of $\mu_1$, $\mu_2$ and $\delta_1$ are related to $\lambda_{{max}}(S^T+S)$. 
	The following two cases are considered. Case 1: $\lambda_{{max}}(S^T+S)>0$. Since $c_4=\max\{\lambda_{max}(S+S^T),c_s\}$, choosing a smaller $c_s$ yields $c_4=\lambda_{{max}}(S^T+S)$. Then, it follows from conditions 1)--2) and 3)--4) that $\mu_1 $ and $\mu_2$ should be selected {to be large enough} to satisfy $\mu_1\!>\!\frac{1}{2\lambda_{min}\!(\mathcal H)}(\frac{\lambda_{max}(S\!+\!S^T)}{p_d-1}+\lambda_{max}(S\!+\!S^T))$ and $\mu_2\!>\!\frac{2\lambda_{max}(S\!+\!S^T)}{(p_d-1)(4\lambda_{min}\!(\bar H))\!^{\frac{a+b}{2b}}}\max\{e^{\frac{c_s(b-a)}{2b}}, $ $e^{1+\frac{\lambda_{max}(S\!+\!S^T)(b-a)p_d\nu_d}{2b(p_d-\!1)}}\}$, respectively.
	It follows from condition 5) that $\delta_1$ should be selected to be larger than $\lambda_{{max}}(S^T+S)$.
	Case 2: $\lambda_{{max}}(S^T+S)\leq0$. 
	Choosing $c_s$ as any positive constant yields $c_4=c_s$. 
	Similarly, $\mu_1 $ and $\mu_2$ should be selected {to be large enough} to satisfy 
	$\mu_1\!>\!\frac{1}{2\lambda_{min}\!(\mathcal H)}(\frac{c_s}{p_d-1}+\lambda_{max}(S\!+\!S^T))$ and $\mu_2\!>\!\frac{2c_s}{(p_d-1)(4\lambda_{min}\!(\bar H))\!^{\frac{a+b}{2b}}}\max\{e^{\frac{c_s(b-a)}{2b}}, $ $e^{1+\frac{c_s(b-a)p_d\nu_d}{2b(p_d-\!1)}}\}$, respectively. And $\delta_1$ can be selected as any positive constant. {Furthermore,} $\mu_3$, $\delta_2$ and $\delta_3$ can be chosen as any positive constants, while $a$ and $b$ are selected to be any positive odd integers with $a<b$.}
	\end{remark}

\begin{remark}
{It follows {from} conditions 2)--4) in Theorem 1 that the {observer} gains $\mu_1$ and $\mu_2$ should be chosen to be larger if the duration of DoS attacks increase in {the} sense that $p_d$ decreases and/or $\nu_d$ increases.}
		\end{remark}

\begin{remark}\label{ef}By letting $\delta_3$ and $\mu_3$ be zero in (\ref{ob}), the proposed distributed resilient fixed-time observer (\ref{ob}) reduces to the distributed resilient finite-time observer.
Moreover, by setting $\delta_2$, $\delta_3$, $\mu_2$ and $\mu_3$ as zero in (\ref{ob}), the proposed distributed resilient fixed-time observer (\ref{ob}) reduces to the distributed resilient {exponentially converging} observer, which is similar to the observer developed in \cite{zhangdananddengchao,li2023111250}.
\end{remark}
\subsection{Distributed Resilient Adaptive Fixed-Time Control Strategy}
Based on the adaptive backstepping technique and the distributed resilient fixed-time observer, a distributed resilient adaptive fixed-time control strategy is developed {in this subsection}.

First, introduce the following coordinate transformations,
\begin{small}\begin{align}\label{v}
	\begin{split}
		z_{i1}&=x_{i1}-X_i\hat v_i,\\
		z_{is}&=x_{is}-X_i\hat v_i^{(s-1)}-\alpha_{i(s-1)},
		s=2, \cdots, n,
	\end{split}
\end{align}\end{small}where $\hat v_i$ is generated by the observer (\ref{ob}),  $X_i^T\!\in\! \mathbb R^q$ {is to} be designed later.

Then, according to the adaptive backstepping technique, the virtual control laws $\alpha_{is}$, $s=1,2,\cdots,n-1$, the final control law $u_i$, and the adaptive law {to estimate} $\theta_i$ {for agent $i$ are designed as follows}:
\begin{small}
	\begin{align}
		\alpha_{i1}\!=\!&-\phi_{i1}^T\!(t,\bar x_{i1})\hat \theta_i\!-\!b_{i1}\hat v_i\!-\!\kappa_{i1}\chi_{i1}^{\frac{3}{2}\!-\!2\beta}z_{i1}^{2\beta\!-\!1}\!-\!\eta_{i1}\!z_{i1}^3\!-\!\rho_{i1}\!z_{i1},\label{controller1}\\
		\alpha_{is}\!=\!&-\!z_{i(s-1)}\!-\!\phi_{is}^T(t,\bar x_{is})\hat \theta_i\!-\!b_{is}\hat v_i\!-\!\kappa_{is}\chi_{is}^{\frac{3}{2}\!-\!2\beta}z_{is}^{2\beta-1}\!-\!\eta_{is}z_{is}^3\nonumber\\
		&-\rho_{is}z_{is}+\sum_{l=1}^{s-1}\!\frac{\partial \alpha_{i(s-1)}}{\partial x_{il}}(x_{i(l+1)}+\phi_{il}^T(t,\bar x_{il})\hat \theta_i+\!b_{il}\hat v_i)\nonumber\\
		&+\sum_{l=1}^{s-1}\!\frac{\partial \alpha_{i(s-1)}}{\partial \hat v_i^{(l-1)}}\hat v_i^{(l)}-\frac{1}{2}(\sum_{l=1}^{s-1}\xi_{il}(\frac{\partial \alpha_{i(s-1)}}{\partial x_{il}})^2)z_{is}\nonumber\\
		&\!+\!\frac{\partial \alpha_{i(s-1)}}{\partial \hat \theta_i}\Gamma_i(\tau_{is}-\zeta_{i1}\hat \theta_i-\zeta_{i2}\hat \theta_i^3)\!+\!(\sum_{l=1}^{s-2}\frac{\partial \alpha_{il}}{\partial \hat\theta_i}z_{i(l+1)})\Gamma_i\nonumber\\
		&\!\times\!(\!\phi_{is}(t,\bar x_{is})\!-\!\!\sum_{l=1}^{s\!-\!1}\!\!\frac{\partial \alpha_{i(s-1)}}{\partial x_{il}}\phi_{il}(t,\!\bar x_{il})), s\!=\!{\!2,3,}\!\cdots\!,n\!-\!1,\label{controllerst}\\
		u_i\!=&-\!z_{i(n-1)}\!+\!X_i\hat v_i^{(n)}\!-\!\phi_{in}^T(t,x_{i})\hat \theta_i\!-\!b_{in}\hat v_i\!-\!\kappa_{in}\chi_{in}^{\frac{3}{2}\!-\!2\beta}z_{in}^{2\beta-1}\nonumber\\
		&-\eta_{in}z_{in}^3-\rho_{in}z_{in}+\sum_{l=1}^{n-1}\!\frac{\partial \alpha_{i(n-1)}}{\partial x_{il}}(x_{i(l+1)}\!+\!\phi_{il}^T(t,\bar x_{il})\hat \theta_i\nonumber\\
		&+b_{il}\hat v_i)+\sum_{l=1}^{n-1}\!\frac{\partial \alpha_{i(n-1)}}{\partial \hat v_i^{(l-1)}}\hat v_i^{(l)}
		\!-\!\frac{1}{2}(\sum_{l=1}^{n-1}\!\xi_{il}(\frac{\partial \alpha_{i(n-1)}}{\partial x_{il}})^2)z_{in}\nonumber\\
		&\!+\frac{\partial \alpha_{i(n-1)}}{\partial \hat \theta_i}\Gamma_i(\tau_{in}\!-\!\zeta_{i1}\hat \theta_i\!-\!\zeta_{i2}\hat \theta_i^3)+(\sum_{l=1}^{n-2}\frac{\partial \alpha_{il}}{\partial \hat \theta_i}z_{i(l+1)})\Gamma_i\nonumber\\
		&\!\times (\phi_{in}(t,x_i)\!-\!\sum_{l=1}^{n-1}\frac{\partial \alpha_{i(n-1)}}{\partial x_{il}}{\phi_{il}(t,\!\bar x_{il})}),\label{controllernt}\\
		\dot{\hat \theta}_i= &\Gamma_i(\tau_{in}-\zeta_{i1}\hat \theta_i-\zeta_{i2}\hat \theta_i^3),\label{adaptive law}
	\end{align}\end{small}where $\beta$, $\zeta_{i1}$, $\zeta_{i2}$, $\kappa_{is}$, $\eta_{is}$, $\rho_{is}$, $\chi_{is}$, $s=1,2,\cdots,n$, $i=1,2,\cdots,$ $N$, are positive constants, $\Gamma_i\!\in\! \mathbb R^{m\times m}$, 
$\xi_{il}=sign(\|b_{il}\|)$, $l=1,2,$ $\cdots,n\!-\!1$, $\hat \theta_i \!\in\! \mathbb R^m$ is the estimate of $\theta_i$, 
$\tau_{is}\!=\!\tau_{i(s-1)}\!+\!(\phi_{is}(t,\bar x_{is})\!-\!\sum_{l=1}^{s-1}\!\frac{\partial \alpha_{il}}{\partial x_{il}}\phi_{il}(t,\bar x_{il}))z_{is}$, $s=2,3,\cdots,n$, with $\tau_{i1}=\phi_{i1}(t,\bar x_{i1})z_{i1}$.

With {the result in} Theorem \ref{lemmat}, the main result of this work can
be given as follows.
\begin{theorem}\label{thm}
{Suppose Assumptions {\ref{assdos}--\ref{assmatrix}} hold.
	Consider the uncertain nonlinear MAS (\ref{agent}) and exosystem (\ref{exo}) under DoS attacks.}
	If conditions 1)--5) in Theorem \ref{lemmat} hold, then the resilient global practical fixed-time COR problem can be solved by the proposed distributed resilient adaptive fixed-time control strategy consisting of the distributed resilient fixed-time observer (\ref{ob}) and the distributed  resilient adaptive fixed-time controller (\ref{controllernt})--(\ref{adaptive law}), where $\Gamma_i$ is a positive definite matrix, $X_i\!=\!-R_i$, $\zeta_{i1}$, $\zeta_{i2}$, $\kappa_{is}$, $\eta_{is}$, $\rho_{is}$, $\chi_{is}$, $s\!=\!1,2,\cdots\!,n$, $i\!=\!1,2,\cdots\!,N$,  are positive constants with $\rho_{is}\!>\!\frac{1}{2}$, {and} $\beta$ is a positive integer.
	Moreover, the residual set is given by
\begin{small}\begin{equation}\label{residual set}
		\left \{e_i: |e_i|\leq {\sqrt{2}}\min \left\{({\frac{NC}{\varpi\eta}})^{\frac{1}{4}},(\frac{C}{\varpi \kappa})^{\frac{2}{3}}
		\right\}\right\},
	\end{equation}\end{small}and the {upper} bound {for} the convergence time {of} the regulated outputs is given by $t_{a}$, 
	where $t_a=t_o+\max\{\frac{4}{\kappa}+\frac{N}{\eta(1-\varpi)}, \frac{4}{\kappa(1-\varpi)}$ $+\frac{N}{\eta}\}$, $t_{o}$ is given by  (\ref{to}), {$\varpi\!\in\! (0,1)$}, $\kappa\!=\!\min_{i=1}^N\{\kappa_i\}$,
$\eta\!=\!\min_{i=1}^N\{\frac{\eta_i}{n+1}\}$, $\kappa_i\!=\!\min\{\!\min_{s=1}^n\{\kappa_{is}\!\}$, $\frac{\zeta_{i1}}{\lambda_{max}^{\frac{3}{4}}(\Gamma_i^{-\!1})}\}$, 
 $\eta_i\!=\!\min\{\!\min_{s=1}^n\{\eta_{is}\},\frac{\zeta_{i2}\min_{{l}=1}^{m}\{4-{9\varepsilon_{il}^{\frac{4}{3}}}\}}{\lambda_{max}^2(\Gamma_i^{-1})}\!\}$,
\begin{small}\begin{equation}\label{C}
		C\!=\!\!\sum_{i=1}^{N}(\sum_{s=1}^{n}\!\kappa_{is}\chi_{is}^{\frac{3}{2}}\!+\!\frac{\zeta_{i1}}{2}\theta_i^T\theta_i\!+\!m\frac{{\zeta_{i1}}}{8}
		\!+\!\zeta_{i2}\!\sum_{l=1}^{m}\!(\frac{1}{12}\!+\!\frac{3}{4\varepsilon_{il}^4})\theta_{il}^4),
	\end{equation}\end{small}$\varepsilon_{il}\in(0, {\frac{2}{3}}^{\frac{3}{2}})$, $l=1,2,\cdots,m$, $i=1,2,\cdots,N$.
\end{theorem}
\textbf{\textit{Proof:}} 
{Define the following Lyapunov function {candidate},}
\begin{small}\begin{align}
	V_{i1}(t)&=\frac{1}{2}z_{i1}^2+\frac{1}{2}\tilde \theta_i^T\Gamma_i^{-1}\tilde \theta_i,\label{V1}\\
	V_{is}(t)&=V_{i(s-1)}(t)+\frac{1}{2}z_{is}^2, s=2,\cdots,n,\label{Vk}\\
	V(t)&=\sum_{i=1}^{N}V_{in}(t),\label{V}
\end{align}\end{small}where $\tilde \theta_i=\theta_i-\hat \theta_i$ denotes the estimation error.

Based on the procedure of the adaptive backstepping technique, the proof can be accomplished in the following steps.

{\textit{Step 1.}} According to (\ref{agent}), (\ref{v}) and (\ref{controller1}), one has
\begin{small}\begin{align}\label{zin1}
	\dot z_{i1}=&z_{i2}+\phi_{i1}^T(t,\bar x_{i1})\tilde \theta_i+b_{i1}(v-\hat v_i)\!-\!\kappa_{i1}\chi_{i1}^{\frac{3}{2}-2\beta}z_{i1}^{2\beta-1}\nonumber\\
	&-\eta_{i1}z_{i1}^3-\rho_{i1}z_{i1}.
\end{align}\end{small}Then, the time derivative of $V_{i1} (t)$ along  (\ref{zin1}) is {given as}
\begin{small}\begin{align}\label{dotvi1}
	\dot{ V}_{i1}(t)\!=&z_{i1}z_{i2}\!+\!\phi_{i1}^T(t,\bar x_{i1})\tilde \theta_iz_{i1}\!+\!b_{i1}(v-\hat v_i)z_{i1}\!-\!\kappa_{i1}\chi_{i1}^{\frac{3}{2}-2\beta}z_{i1}^{2\beta}\nonumber\\
	&-\eta_{i1}z_{i1}^4-\rho_{i1}z_{i1}^2-\tilde \theta_i^T\Gamma_i^{-1}\dot{\hat \theta}_i.
\end{align}\end{small}For term $-\kappa_{i1}\chi_{i1}^{\frac{3}{2}-2\beta}z_{i1}^{2\beta}$ in (\ref{dotvi1}), one has
\begin{small}\begin{align*}
	-\kappa_{i1}\chi_{i1}^{\frac{3}{2}-2\beta}z_{i1}^{2\beta}\leq\left\{
	\begin{array}{l}
		-\kappa_{i1}(z_{i1}^2)^{\frac{3}{4}}, \chi_{i1}< |z_{i1}|,\\
		-\kappa_{i1}(z_{i1}^2)^{\frac{3}{4}}+\kappa_{i1}\chi_{i1}^{\frac{3}{2}}, 
		\chi_{i1}\geq |z_{i1}|.
	\end{array}
	\right.
\end{align*}\end{small}By Lemma \ref{lemma1}, one further has 
\begin{small}\begin{align}\label{zi1case1}
	\dot{ V}_{i1}(t)\leq&z_{i1}z_{i2}+\frac{1}{2}\|b_{i1}(v-\hat v_i)\|^2-\kappa_{i1}(z_{i1}^2)^{\frac{3}{4}}-\eta_{i1}(z_{i1}^2)^2\nonumber\\
	&-(\rho_{i1}-\frac{1}{2})z_{i1}^2+\tilde \theta_i^T(\tau_{i1}-\Gamma_i^{-1}\dot{\hat \theta}_i)+\kappa_{i1}\chi_{i1}^{\frac{3}{2}}.
\end{align}\end{small}\quad {\textit{Step $s$ $({s=2,3},\cdots, n-1)$.}} 
By (\ref{agent}), (\ref{v}) and (\ref{controllerst}), one has
\begin{small}\begin{align}\label{zins}
\dot z_{is}\!\!=&z_{i(s\!+\!1)}\!\!-\!\!z_{i(s\!-\!1)}\!\!+\!\phi_{is}^T(\!t,\bar x_{is}\!)\tilde \theta_i\!+\!b_{is}\!(v\!-\!\hat v_i)\!-\!\kappa_{is}\!\chi_{is}^{\frac{3}{2}\!-\!2\beta}\!z_{is}^{2\beta\!-\!1}\!\!-\!\eta_{is}\!z_{is}^3\nonumber\\
	&\!-\!\rho_{is}z_{is}\!-\!\sum_{l=1}^{s-1}\!\!\frac{\partial \alpha_{i(s-1)}}{\partial x_{il}}\phi_{il}^T(t,\bar x_{il})\tilde \theta_i\!+\!\sum_{l=1}^{s-1}\!\!\frac{\partial \alpha_{i(s-1)}}{\partial x_{il}}b_{il}(\hat v_i-v)\nonumber\\
	&\!\!+\!\frac{\partial \alpha_{i(s\!-\!1)}}{\partial \hat \theta_i}\!\Gamma_i\!(\!\tau_{is}\!-\!\zeta_{i1}\hat \theta_i\!-\!\zeta_{i2}\hat \theta_i^3\!\!-\!\Gamma_i^{-1}\!\dot{\hat \theta}_i\!)\!-\!\frac{1}{2}\!(\sum_{l=1}^{s\!-\!1}\!\xi_{il}(
	\!\frac{\partial \alpha_{i(s\!-\!1)}}{\partial x_{il}}\!)\!^2)z_{is}\nonumber\\
	&\!+\!(\!\sum_{l=1}^{s-2}\!\frac{\partial \alpha_{il}}{\partial \hat\theta_i}z_{i(l+1)}\!)\Gamma_i\!(\!\phi_{is}(t,\bar x_{is}\!)\!-\!\!\sum_{l=1}^{s-1}\!\!\frac{\partial \alpha_{i(s\!-\!1)}}{\partial x_{il}}\phi_{il}(t,\bar x_{il})\!).
\end{align}\end{small}Then, together with (\ref{zi1case1}), taking the derivative of $V_{is} (t)$ along (\ref{zins}) leads to
\begin{small}\begin{align}\label{ziscaseall}
	\dot V_{is}(t)\!\!\leq&z_{is}z_{i(s+1)}-\sum_{l=1}^{s}	\kappa_{il}(z_{il}^2)^{\frac{3}{4}}-\sum_{l=1}^{s}\eta_{il}(z_{il}^2)^2-\sum_{l=1}^{s}(\rho_{il}-\frac{1}{2})z_{il}^2\nonumber\\
	&+\sum_{l=0}^{s-1}\frac{s-l}{2}\|b_{i(l+1)}(\hat v_i\!-\!v)\|^2\!+\!\tilde \theta_i^T\!(\tau_{is}\!-\!\Gamma_i^{-1}\dot{\hat \theta}_i)\!+\!\sum_{l=1}^{s}\!\kappa_{il}\chi_{is}\!^{\frac{3}{2}}\nonumber\\
	&+(\sum_{l=1}^{s-1}\!\frac{\partial \alpha_{il}}{\partial{\hat \theta_i}}z_{i(l+1)})\Gamma_i(\tau_{is}\!-\!\zeta_{i1}\hat \theta_i\!-\!\zeta_{i2}\hat \theta_i^3\!-\!\Gamma_i^{-1}\dot{\hat \theta}_i).
\end{align}\end{small}\quad {\textit{Step $n$.}} 
According to (\ref{agent}), (\ref{v}) and (\ref{controllernt}), one can obtain that
\begin{small}\begin{align}\label{zint}
	\dot z_{in}\!=\!&\!-\!z_{i(n-1)}\!+\!\phi_{in}^T(t,x_{i})\tilde \theta_i\!+\!b_{in}(v\!-\!\hat v_i)\!-\!\kappa_{in}\chi_{in}^{\frac{3}{2}\!-\!2\beta}z_{in}^{2\beta\!-\!1}\!\!-\!\eta_{in}z_{in}^3\nonumber\\
	&-\rho_{in}z_{in}\!-\!\sum_{l=1}^{n-1}\!\!\frac{\partial \alpha_{i(n-1)}}{\partial x_{il}}\phi_{il}^T(t,\bar x_{il})\tilde \theta_i\!+\!\sum_{l=1}^{n-1}\!\frac{\partial \alpha_{i(n-1)}}{\partial x_{il}}b_{il}(\hat v_i\!-\!v)\nonumber\\
	&\!\!+\!\frac{\partial \alpha_{i(n\!-\!1)}}{\partial \hat \theta_i}\!\Gamma_i\!(\tau_{in}\!\!-\!\zeta_{i1}\hat \theta_i\!-\!\zeta_{i2}\hat \theta_i^3\!\!-\!\Gamma_i^{-1}\!\dot{\hat \theta}_i)\!-\!\!\frac{1}{2}\!(\!\sum_{l=1}^{n\!-\!1}\!\!\xi_{il}\!(\!\frac{\partial \alpha_{i(n\!-\!1)}}{\partial x_{il}}\!)\!^2)z_{in}\nonumber\\
	&\!+\!(\!\sum_{l=1}^{n\!-\!2}\!\frac{\partial \alpha_{il}}{\partial \hat \theta_i}z_{i(l+1)}\!)\Gamma_i\!(\!\phi_{in}(t,x_i)\!-\!\!\sum_{l=1}^{n-1}\!\!\frac{\partial \alpha_{i(n-1)}}{\partial x_{il}}\phi_{il}(\!t,\!\bar x_{il}\!)\!).
\end{align}\end{small}{Noting} (\ref{ziscaseall}), the time derivative of $V_{in} (t)$ along (\ref{zint}) is {given as}
\begin{small}\begin{align}\label{ziscaneall}
	\dot V_{in}(t)\leq& -\sum_{l=1}^{n}\kappa_{il}(z_{il}^2)^{\frac{3}{4}}-\sum_{l=1}^{n}\eta_{il}z_{il}^{4}-\sum_{l=1}^{n}	(\rho_{il}-\frac{1}{2})z_{il}^{2}\nonumber\\
	&
	+\tilde \theta_i^T(\tau_{in}-\Gamma_i^{-1}\dot{\hat \theta}_i)+(\sum_{l=1}^{n-1}\frac{\partial \alpha_{il}}{\partial{\hat \theta_{i}}}z_{i(l+1)})\Gamma_i(\tau_{in}\nonumber\\
	&-\zeta_{i1}\hat \theta_i-\zeta_{i2}\hat \theta_i^3
	-\Gamma_i^{-1}\dot{\hat \theta}_i)+\sum_{l=1}^{n}\kappa_{il}\chi_{is}^{\frac{3}{2}}+\Theta_i(t),
\end{align}\end{small}where $\Theta_i\!=\sum_{l=0}^{n-1}\!\frac{n-l}{2}\!\|b_{i(l+1)}(\hat v_i-v)\|^2$.
Since $\rho_{il}>\frac{1}{2}$, it follows from (\ref{adaptive law}) and (\ref{ziscaneall}) that 
\begin{small}\begin{align}\label{ziscaneall2}
	\dot V_{in}(t)\leq& -\sum_{l=1}^{n}	\kappa_{il}(z_{il}^2)^{\frac{3}{4}}-\sum_{l=1}^{n}\eta_{il}(z_{il}^2)^{2}+\zeta_{i1}\tilde \theta_i^T\hat \theta_i\nonumber\\
	&+\zeta_{i2}\tilde \theta_i^T\hat \theta_i^3
	+\sum_{l=1}^{n}\kappa_{il}\chi_{is}^{\frac{3}{2}}+\Theta_i.
\end{align}\end{small}{{By Lemma \ref{lemma1}, one has}
		\begin{small}\begin{subequations}\label{label1}
				\begin{align}
\tilde \theta_i^T\hat \theta_i
	\!\leq\!&\frac{1}{2}\theta_i^T\theta_i+\frac{m}{8}
	\!-\!\frac{1}{\lambda_{max}^{\frac{3}{4}}(\Gamma_i^{-1})}(\frac{\tilde \theta_i^T\Gamma_i^{-1}\tilde \theta_i}{2})^{\frac{3}{4}},\\
\tilde \theta_i^T\hat \theta_i^3
\leq&\!\sum_{l\!=\!1}^{m}\!(\!\frac{1}{12}\!+\!\frac{3}{4\varepsilon_{il}^4}\!)\theta_{il}^4
\!-\!\frac{\min_{l=1}^m(4-{9\varepsilon_{il}^{\frac{4}{3}}})}{\lambda_{max}^2\!(\Gamma_i^{-1}\!)}\!(\!\frac{\tilde \theta^T_i\Gamma_i^{-1}\!\tilde \theta_i}{2})^2,
\end{align}\end{subequations}\end{small}where $0<\varepsilon_{il}<{\frac{2}{3}}^{\frac{3}{2}}$.}

Substituting (\ref{label1}) into (\ref{ziscaneall2}) leads to
\begin{small}\begin{align}\label{ziscaneall2.1}
\dot V_{in}(t)\!\leq&\! -\!\sum_{l=1}^{n}\!	\kappa_{il}(z_{il}^2)^{\frac{3}{4}}\!-\!\sum_{l=1}^{n}\!\eta_{il}(z_{il}^{2})^2\!-\!\frac{\zeta_{i1}}{\lambda_{max}^{\frac{3}{4}}(\Gamma_i^{\!-\!1})}\!(\!\frac{\tilde \theta_i^T\Gamma_i^{-1}\tilde \theta_i}{2}\!)^{\frac{3}{4}}\nonumber\\
&-\frac{\zeta_{i2}\min_{l=1}^m(4-{9\varepsilon_{il}^{\frac{4}{3}}})}{\lambda_{max}^2(\Gamma_i^{\!-\!1})}(\frac{\tilde \theta^T_i\Gamma_i^{\!-\!1}\tilde \theta_i}{2})^2+\Theta_i
+c_{ni},
\end{align}\end{small}where $c_{ni}=\sum_{l=1}^{n}\kappa_{il}\chi_{is}^{\frac{3}{2}}+\frac{\zeta_{i1}}{2}\theta_i^T\theta_i+m\frac{{\zeta_{i1}}}{8}+\zeta_{i2}\sum_{l=1}^{m}(\frac{1}{12}+\frac{3}{4\varepsilon_{il}^4})\theta_{il}^4$.
{It follows from
(\ref{V}), (\ref{ziscaneall2.1}), Lemma \ref{le1}, and the definitions of $\kappa$ and $\eta$ that} 
\begin{small}\begin{align}\label{all}
\dot V(t)
&\leq -\kappa (V(t))^{\frac{3}{4}}-\frac{\eta}{N}(V(t))^{2}+\sum_{i=1}^{N}\Theta_i(t)+C,
\end{align}\end{small}where $C$ is given by (\ref{C}).
From Theorem \ref{lemmat}, one can infer that 
{there is} a positive constant $c_v$ such that $\sum_{i=1}^{N}\Theta_i\leq c_v$,  for any $v(0)$, $\hat v_i(0)\in \mathbb R^q$.
Then,  when $(V(t))^2\geq \frac{N(C+c_v)}{\eta}$, one has {
	$\dot V(t) \leq 0.$}
{It then follows from the standard analysis as in \cite{krstic1995nonlinear} that} all the signals in the {resulting} closed-loop system are globally bounded.

 Since for any $v(0)$, $\hat v_i(0)\in \mathbb R^q$, $\hat v_i-v=0$, $\forall t\geq t_o$ from Theorem \ref{lemmat}, one has $\sum_{i=1}^N\Theta_i=0$, $\forall t\geq t_o$. By (\ref{all}), one further has
 \begin{small}\begin{align}\label{all1}
 	\dot V(t)
 	&\leq -\kappa (V(t))^{\frac{3}{4}}-\frac{\eta}{N}(V(t))^2+C, \forall t\geq t_o.
 \end{align}\end{small}To show the global fixed-time convergence of the regulated outputs, the following two cases are considered.

\textit{Case 1:} $(V(t))^2\geq \frac{NC}{\varpi \eta}$.

{In this case, it follows from (\ref{all1}) that}
\begin{small}\begin{align}\label{all2}
\dot V(t)
&\leq -\kappa (V(t))^{\frac{3}{4}}-\frac{\eta}{N}(1-\varpi)(V(t))^2, \forall t\geq t_o.
\end{align}\end{small}Then, by Lemma \ref{fixed time stability}, one can infer that $V(t)$ globally converges to the residual set $\{V(t): V(t)\leq ({\frac{NC}{\varpi \eta}})^{\frac{1}{2}}\}$ in a fixed time, where its upper bound is given by $t_{a1}=t_o+\frac{4}{\kappa}+\frac{N}{\eta(1-\varpi)}$.

\textit{Case 2:}
$(V(t))^{\frac{3}{4}}\geq \frac{C}{\varpi \kappa}$.

{Similarly in this case, it} follows from (\ref{all1}) and Lemma \ref{fixed time stability} that $V(t)$ globally converges to the residual set $\{V(t): V(t)\leq (\frac{C}{\varpi \kappa})^{\frac{4}{3}}\}$ in a fixed time with its upper bound giving by  $t_{a2}=t_o+\frac{4}{\kappa(1-\varpi)}+\frac{N}{\eta}$.

{Combining the} results obtained in {the} above two cases, one can infer that $z_{i1}$ globally converges to the residual set $\left \{z_{i1}: |z_{i1}|\leq{\sqrt{2}}\min \left\{({\frac{NC}{\varpi\eta}})^{\frac{1}{4}},(\frac{C}{\varpi \kappa})^{\frac{2}{3}}
\right\}\right\}$ in a fixed time $t_a=\max\{t_{a1},t_{a2}\}$. Since for any $v(0)$, $\hat v_i(0)\in \mathbb R^q$, $\hat v_i-v=0$, $\forall t\ge t_{o}$ from Theorem \ref{lemmat}, $X_i=-R_i$ and $z_{i1}=x_{i1}-X_i \hat v_i$, {thus one concludes that} the regulated output $e_i$ globally converges to the residual set in a fixed time $t_a$, where the residual set is defined as in (\ref{residual set}).
This completes the proof.

\begin{remark}\label{sin}
$\chi_{is}$ and $\beta$ in terms $-\kappa_{is}\chi_{is}^{\frac{3}{2}-2\beta}z_{is}^{2\beta-1}$, $s=1,2,\cdots,n$, are introduced to handle the singularity problem.
\end{remark}
\begin{remark}
In Theorem \ref{thm}, the residual set (\ref{residual set}) is tunable {in terms of those design parameters,} {$\kappa_{is}$, $\eta_{is}$, $\chi_{is}$, $\zeta_{i1}$, $\zeta_{i2}$ and ${\Gamma_i}$, $s\!=\!1,2,\cdots\!,n$, $i\!=\!1,2,\cdots\!,N$.} {It follows from  (\ref{residual set}) that} larger $\kappa$, $\eta$ and/or smaller $C$ yield a smaller  residual set.
{One can obtain a smaller $C$ by decreasing $\chi_{is}$, $\zeta_{i1}$ and $\zeta_{i2}$ such that $\kappa_{is}\chi_{is}^{\frac{3}{2}}$, $\frac{\zeta_{i1}}{2}\theta_i^T\theta_i$, $m\frac{\zeta_{i1}}{8}$ and $\zeta_{i2}\sum_{l=1}^{m}(\frac{1}{12}\!+\!\frac{3}{4\varepsilon_{il}^4})\theta_{il}^4$ are decreased.
Moreover, one can obtain larger $\kappa$ and $\eta$ by increasing $\kappa_{is}$, $\eta_{is}$ and $\lambda_{max}\!(\Gamma_i)$ such that $\kappa_i$, $\eta_i$, 
 $\frac{\zeta_{i1}}{\lambda_{max}^{\frac{3}{4}}\!(\Gamma_i^{-1})}$ and $\frac{\zeta_{i2}\min_{l=1}^{m}\{4-{9\varepsilon_{il}^{\frac{4}{3}}}\}}{\lambda_{max}^2\!(\Gamma_i^{-1})}$ are increased.}
\end{remark}
\section{A Simulation Example}
In this section, an example is provided to demonstrate the efficacy of the proposed distributed resilient adaptive fixed-time control strategy.
Consider a MAS consisting of four agents.
Each agent is a one-link manipulator with the inclusion of motor
dynamics {borrowed from \cite{chen2009adaptive}}, whose dynamics are described by 
\begin{small}\begin{eqnarray}\label{sagent}
	\begin{array}{lll}
		&D_i\ddot \omega_i+B_i \dot \omega_i+N_i\sin(\omega_i)=\varrho_i+\varrho_{id},\\
		&M_i\dot \varrho_i+G_i\varrho_i=u_i-K_{i}\dot \omega_i,
	\end{array}
\end{eqnarray}\end{small}{where $\omega_i$, $\dot \omega_i$,  $\ddot \omega_i$ denote {the} link {angular} position, velocity,  and acceleration, respectively, $\varrho_i$ and $\dot \varrho_i$ represent the motor shaft angle and velocity{, respectively,}
$u_i$ is {the} control input representing the motor torque, {and} $\varrho_{id}$ represents the torque disturbance.}
 The torque disturbance $\varrho_{id}={[0, 1]}v$, {where $v\in \mathbb R^2$ is} the exogeneous signal generated by {exosystem} (\ref{exo}) with $S=
\left[\begin{smallmatrix}
	\setlength{\arraycolsep}{0.1pt}
	0&1\\
	-1&0\\
\end{smallmatrix}\right].$ 

\begin{figure}[htbp]
			\centering
				\vspace{-0.3cm} 
			\begin{minipage}[b]{0.5\textwidth} 
				\centering 
				\includegraphics[height=0.7cm,width=3.6cm]{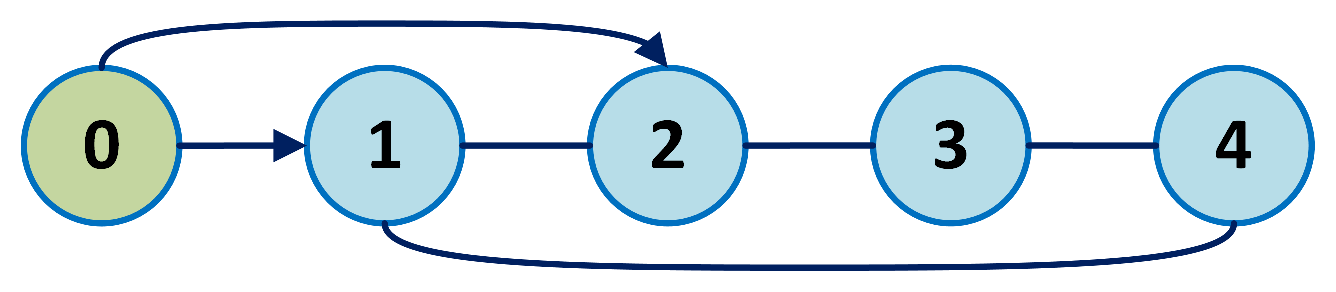}
	            \caption{\label{fig1}Communication graph $\mathcal {\bar G}$.}
			\end{minipage}
			\vspace{-0.3cm} 
		\end{figure}

{The parameters of the agents, along with their appropriate units provided in \cite{chen2009adaptive}, are {given} as follows:}
$D_i=1$, $M_i=1$, $B_i=0.2\cdot i$, $N_i=0.1\cdot i$,  $K_{i}=0.2\cdot i$, $G_i=0.1\cdot i$.
Let $x_{i1}=\omega_i$, $x_{i2}=\dot \omega_i$, $x_{i3}=\varrho_i$. Then, system (\ref{sagent}) can be expressed 
as in (\ref{agent})
%
with $b_{i1}=b_{i3}=[0, 0]$, $b_{i2}=[0, 1]$, $\phi_{i1}^T(t,\bar x_{i1})=[0, 0, 0, 0]$, $\phi_{i2}^T(t,\bar x_{i2})=[\sin(x_{i1}), x_{i2}, 0, 0]$, $\phi_{i3}^T( t,x_{i})=[0, 0, x_{i2}, x_{i3}]$, $\theta_{i}^T=[-0.1\cdot i,-0.2\cdot i,-0.2\cdot i,-0.1\cdot i]$, and $R=[1, 0]$.
Note that $\theta_i$ is supposed to be unknown in this study.
%
The communication topology among {the} agents and exosystem is given in Figure \ref{fig1}, {and} the {corresponding matrix} $\mathcal H$ is given as $\mathcal H=\left[\begin{smallmatrix}
			\renewcommand\arraystretch{0.5}
			\setlength{\arraycolsep}{0.2pt}
			3&-1&0&-1\\
			-1&3&-1&0\\
			0&-1&2&-1\\
			-1&0&-1&2
		\end{smallmatrix}\right].$

		\begin{table}[htbp]
			\centering
			\setlength{\abovecaptionskip}{-0.05cm} 
			\setlength{\abovedisplayskip}{-2cm} 
			\setlength{\belowdisplayskip}{-0.5cm} 
			\caption{ DoS attacks intervals}
		\scalebox{0.8}{\begin{tabular}{|l|l|}
					\hline
					Edge& DoS attacks intervals \\
					\hline
					(1,0) & $[0.55,1.2)\bigcup[1.95,2.54)\bigcup[3.65,4.5)$ \\
					\hline
					(2,0) & $[0.01,0.5)\bigcup[1.3,1.9)\bigcup[4.8,5.2)\bigcup[8,8.2)$ \\
					\hline
					(2,1) & $[0.01,0.5)\bigcup[0.55,1.2)\bigcup[1.3,1.9)\bigcup[1.95,2.54)\bigcup[3.65,4.5)$\\
					&$\bigcup[4.8,5.2)\bigcup[8,8.2)$ \\
					\hline
					(3,2)&$[0.01,0.5)\bigcup[0.55,1.2)\bigcup[1.3,1.9)\bigcup[1.95,2.54)\bigcup[2.7,3.64)$\\
					&$\bigcup[3.65,4.5)\bigcup[4.8,5.2)\bigcup[8,8.2)$ \\
					\hline
					(4,3)&$[0.01,0.5)\bigcup[0.55,1.2)\bigcup[1.3,1.9)\bigcup[1.95,2.54)\bigcup[2.7,3.64)$\\
					&$\bigcup[3.65,4.5)\bigcup[4.8,5.2)\bigcup[7.7,7.9)\bigcup[8,8.2)$ \\
					\hline
					(1,4)&$[0.01,0.5)\bigcup[0.55,1.2)\bigcup[1.3,1.9)\bigcup[1.95,2.54)\bigcup[3.65,4.5)$\\
					&$\bigcup[4.8,5.2)\bigcup[7.7,7.9)\bigcup[8,8.2)$ \\
					\hline
			\end{tabular}}
		\end{table}
		\begin{figure}[htbp]
		\vspace{-0.2cm}  
			\centering
			\subfigure[]
			{\includegraphics[height=3.1cm,width=4.2cm]{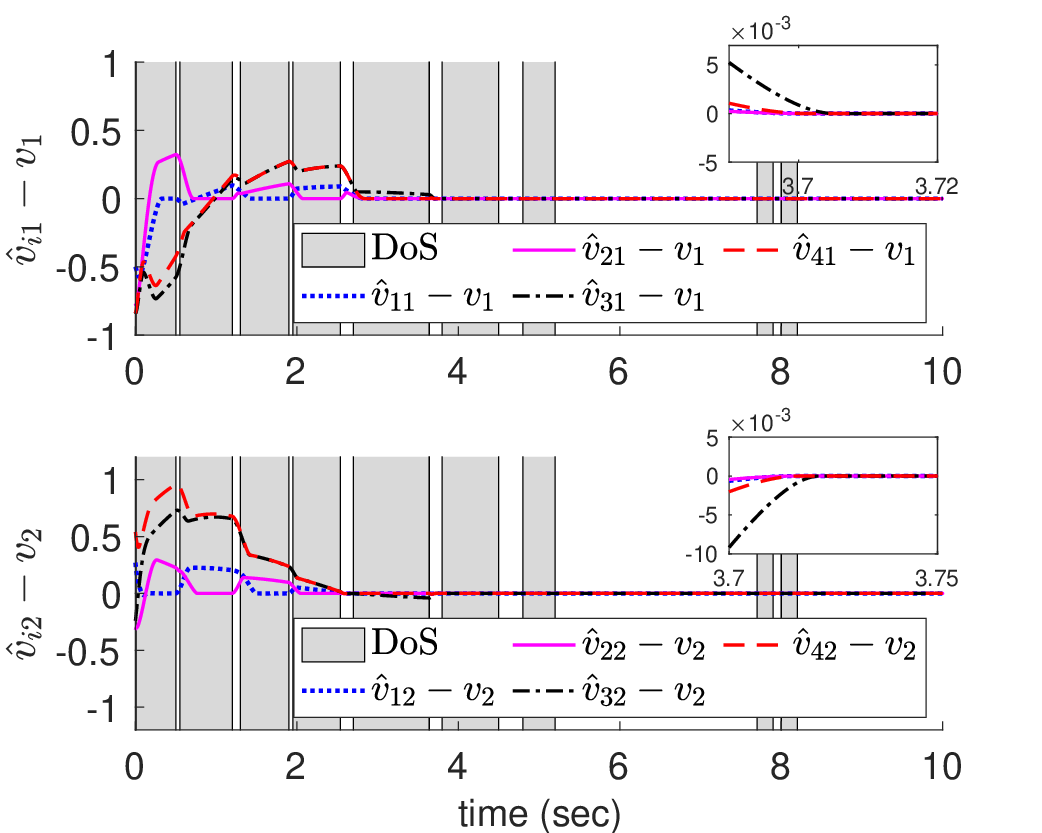}}
			\hspace{-0.5cm}
			\subfigure[]
			{\includegraphics[height=3.1cm,width=4.2cm]{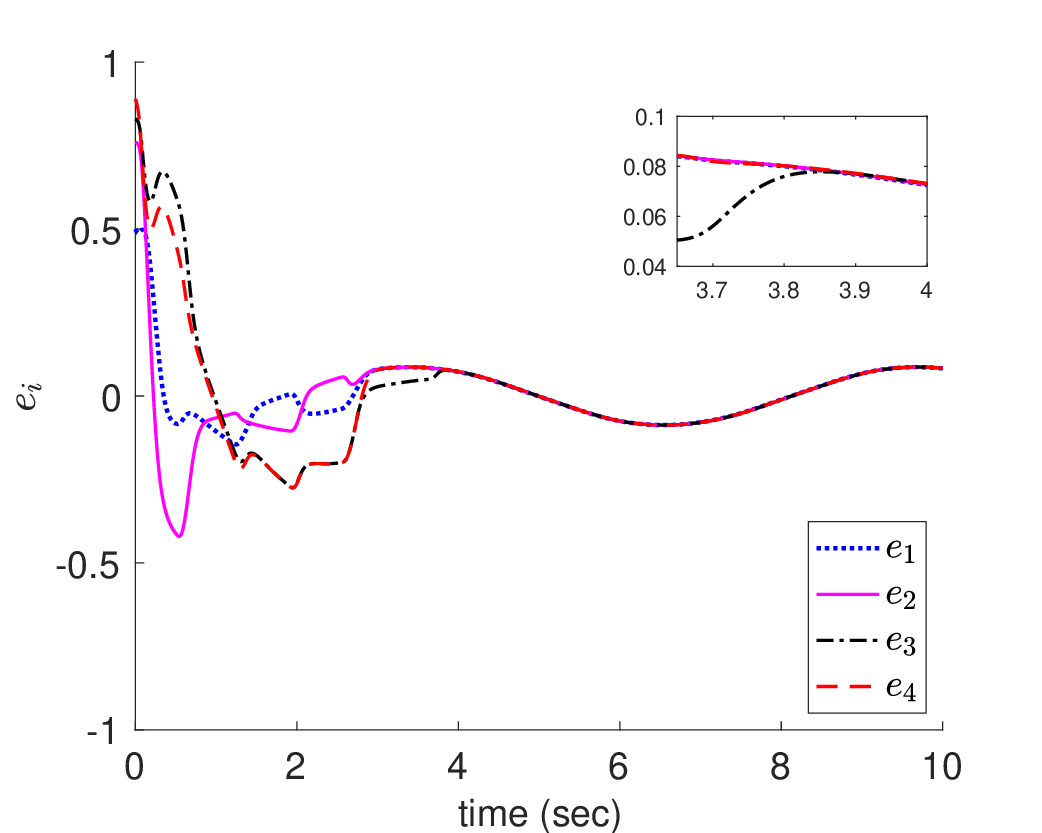}}
	\vspace{-0.3cm} 
		\caption{{Responses of the estimation errors and regulated outputs.}} 
		\label{fig2}
	\end{figure}	
	\begin{figure}[htbp]
	\vspace{-0.5cm} 
			\centering
		\subfigure[]
			{\includegraphics[height=3.7cm,width=4.4cm]{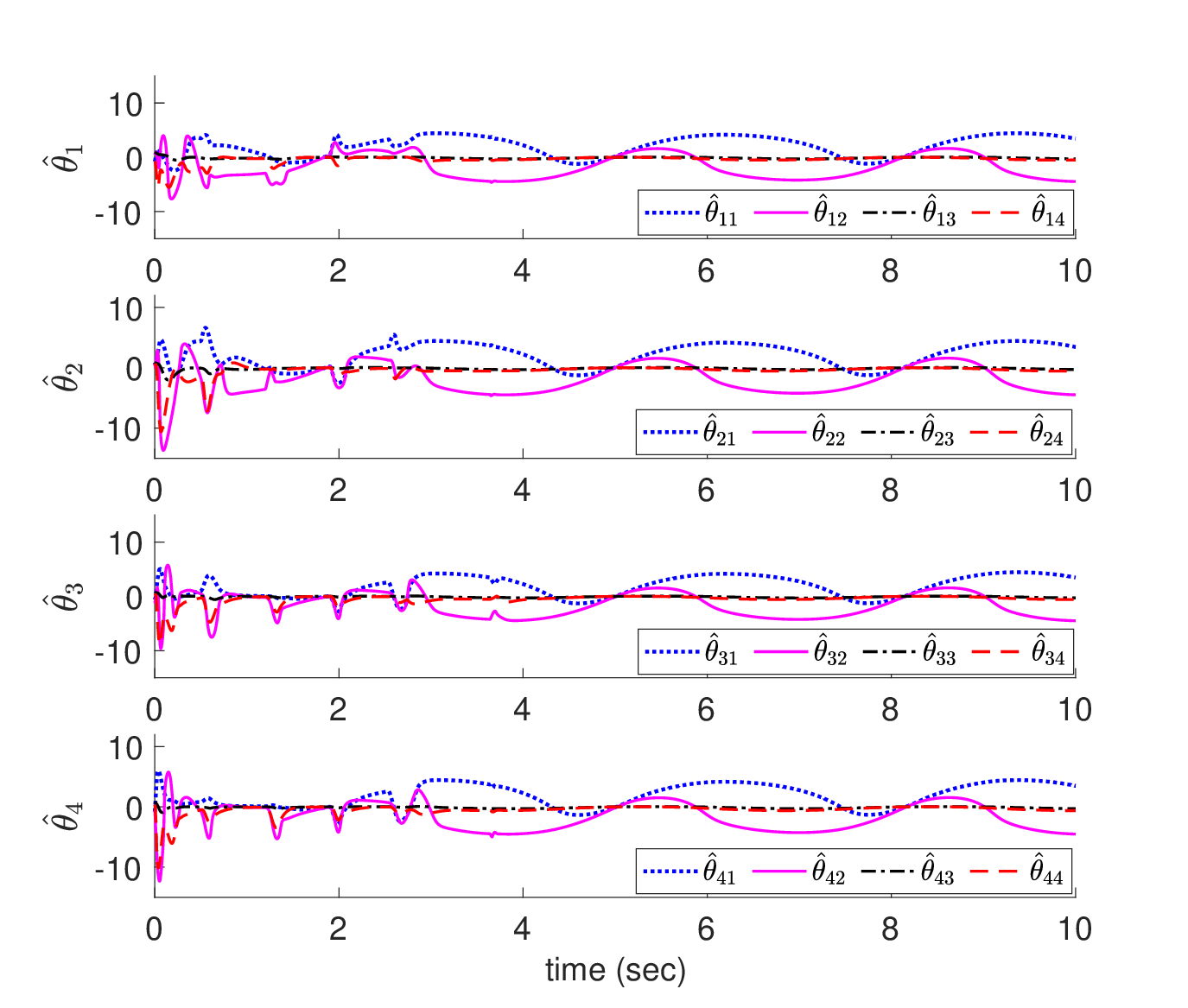}}
				\hspace{-0.5cm}
					\subfigure[]
			{\includegraphics[height=3.5cm,width=4.4cm]{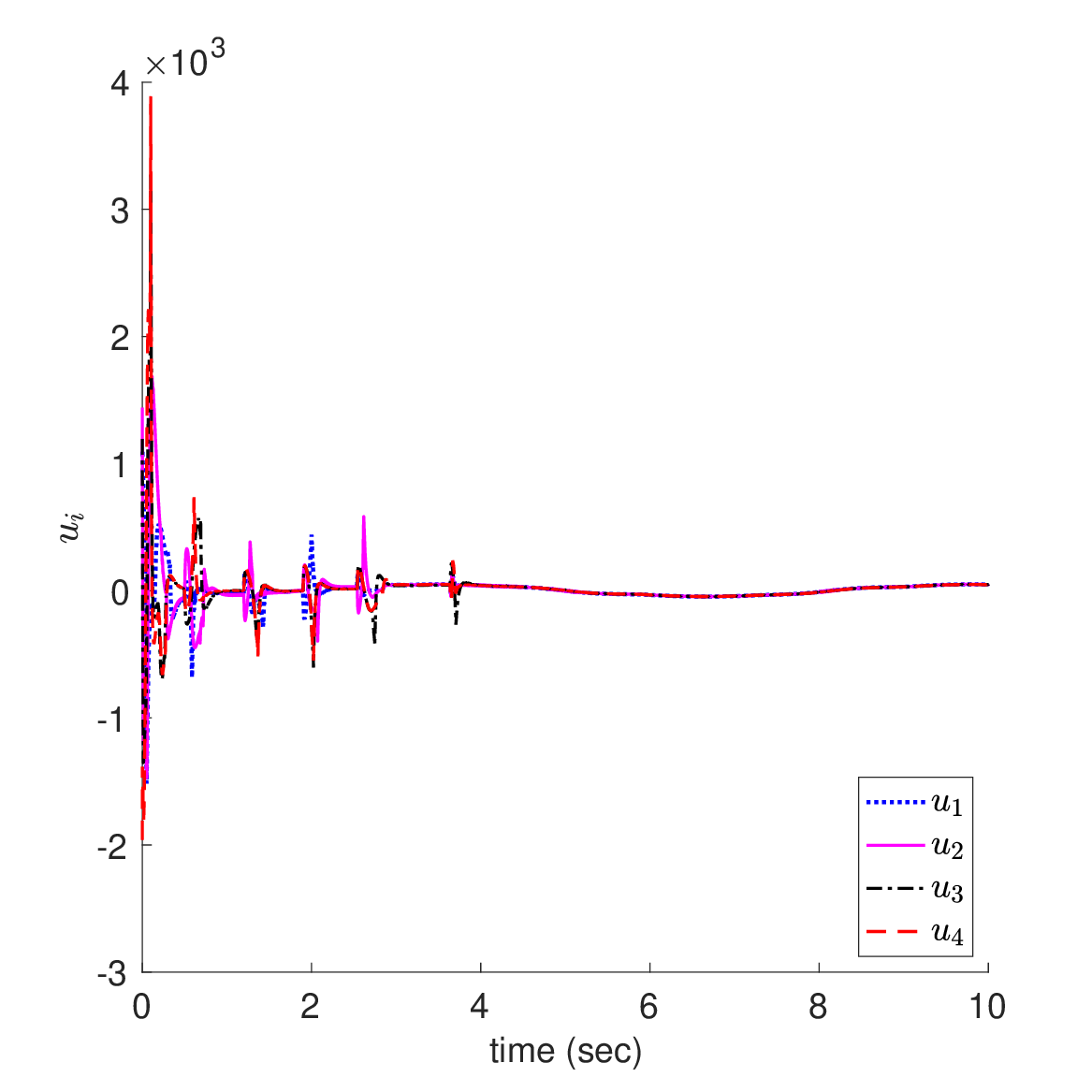}}
	\vspace{-0.3cm} 
	\caption{\label{fig4}{{Evolutions} of the estimated parameters and control {signals}.}}
\	\setlength{\belowcaptionskip}{-0.5cm}
	\end{figure}
		Set the DoS attack parameters as $p_d=4$ {and} $\nu_d=0.1$.
		Table I shows the DoS attacks intervals over $t\in[0,20)$ for edges {$(i,j)\in \bar {\mathcal E}$}.
		One can calculate that $\lambda_{min}(\mathcal H)=\lambda_{min}(\mathcal {\bar H})=\lambda_{min}(\mathcal{\tilde H})=0.3820$ and $\lambda_{{max}}(S+S^T)=0$.	
		Select the parameters of the {distributed} observer as $\mu_1=1.6$, {$\mu_2=2.1$}, $\mu_3=4.1$, $\delta_1=4$, $\delta_2=6$, $\delta_3=6$, $a=3$, $b=7$. {Then, one can calculate that $c_1=1.2223$, $c_2=4.6290$ and $c_3=1.0774$.
		Choose $c_s=2$, one has $c_4=2$.}  Then, one can verify that conditions 1)--5) in Theorem \ref{lemmat} are satisfied. It follows from (\ref{to}) that the upper bound for the settling time of {the  distributed observer} is $t_{o}=24.9113$ sec.
	    Select the parameters of the virtual control laws $\alpha_{i1}(t)$, $\alpha_{i2}(t)$, the final control law $u_i(t)$ and the adaptive law as follows: $\beta=1$, $\chi_{il}=0.01$, $l=1,2,3$,  {$\zeta_{i1}=1$,} $\zeta_{i2}=0.1$, $\Gamma_i=diag(11, 11, 11, 11)$, {$\kappa_{is}=2$, $\eta_{is}=1.5$, $\rho_{is}=2$,} $s=1,2,3$. Choose $\varepsilon_{il}=0.5$, $i, l=1,\cdots,4$, ${\varpi}=0.5$. Then, one can calculate that $\kappa=2$, $\eta=0.3750$, $C= 3.6784$ and $t_a=48.2446$ sec.

{Several simulations have been conducted, where the initial condition{s are} randomly generated in set $(-1,1)$. One of {the} simulation results is recorded {in Figures \ref{fig2}-\ref{fig4}}, where {$v(0)\!=\![0.62, -0.40]^T\!$, $\hat v_1(0)\!=\![0.12,-0.13]^T\!$, $\hat v_2(0)\!=\![-0.15,0.10]^T\!$, $\hat v_3(0)\!=\![-0.20, 0.54]^T\!$, $\hat v_4(0)\!=\![0.16, 0.85]^T$, 
$x_1(0)\!=\![ -0.16, -0.28, 0.11]^T\!$, $x_2(0)\!=\![ 0.19, 0.46, 0.14]^T\!$, $x_3(0)\!=\![0.15, 0.25, 0.32]^T\!$, and $x_4(0)\!=\![0.29, 0.15, -0.23]^T\!$.}}
	From Figure \ref{fig2} {(a)}, one can observe that the estimation errors converge to zero in {a} fixed time.
	It can be seen from Figure \ref{fig2} {(b)} that the regulated outputs converge to a residual set in {a} fixed time under the proposed control strategy. {It means that the manipulator's  link angular position $\omega_i$ can practically converge to the signal $-R_iv$ in a fixed time.} Besides, Figure \ref{fig4} (a) verifies the boundedness of the estimated parameters for the agents, {while Figure \ref{fig4} (b) demonstrates the evolution of the control {signals}.}

		\begin{figure}[htbp]
	\centering
	\begin{minipage}[htbp]{0.24\textwidth} 
		\centering
		\setlength{\abovecaptionskip}{-0.1cm} 	
		\setlength{\belowcaptionskip}{-0.1cm}  
		\includegraphics[height=1.7cm,width=4.4cm]{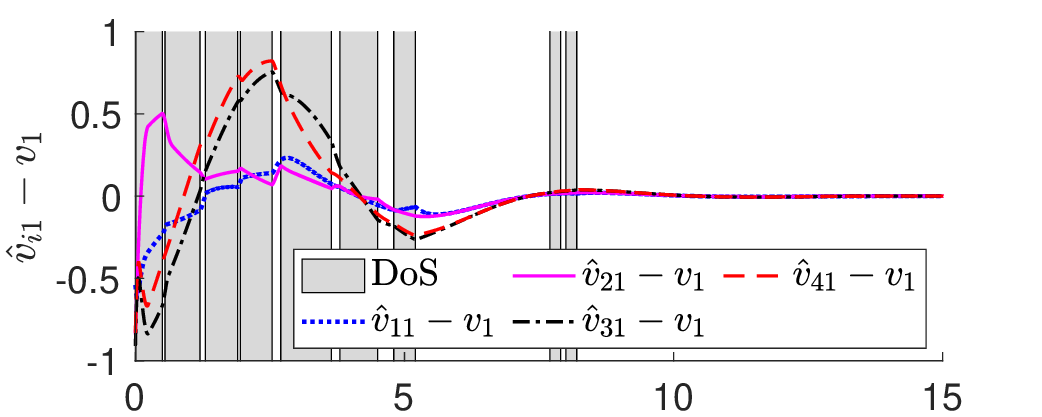}
	\end{minipage}
	\begin{minipage}[htbp]{0.24\textwidth} 
		\centering
		\setlength{\abovecaptionskip}{-0.1cm} 	
		\setlength{\belowcaptionskip}{-0.1cm} 
		\includegraphics[height=1.7cm,width=4.4cm]{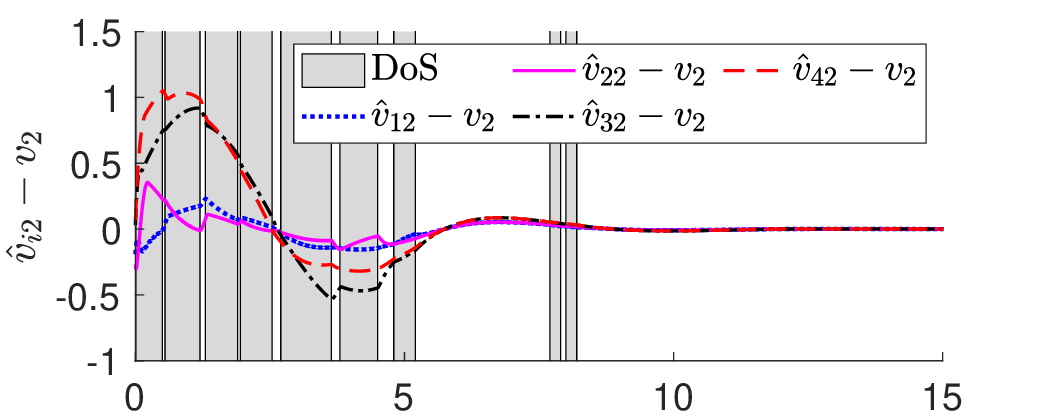}
	\end{minipage}
	\caption{\label{fig2e}{Responses of the estimation errors for the distributed resilient {exponentially converging} observer developed in \citep{li2023111250}.}}
\end{figure}	
%
	
{Besides, a comparison of the convergence performance among the distributed resilient fixed-time observer (\ref{ob}) and the distributed resilient {exponentially converging} observer developed in \citep{li2023111250} has been conducted}. 
{The responses of the estimation errors for the distributed resilient {exponentially converging} observer developed in \citep{li2023111250} is shown in Figure \ref{fig2e}. One can observe from Figures \ref{fig2} {(a)} and  \ref{fig2e} that the distributed resilient fixed-time observer (\ref{ob}) has a faster convergence speed than the distributed resilient {exponentially converging} observer developed in \citep{li2023111250}.}	
\section{Conclusions}
This paper develops a novel distributed resilient adaptive fixed-time control strategy to address the resilient global practical fixed-time COR problem of uncertain nonlinear MASs under DoS attacks. 
{It is shown that all the signals in the resulting closed-loop system are globally bounded and the regulated outputs converge to a {tunable} residual set within a fixed time. Furthermore, the upper bound of the convergence time is explicitly given. The guidelines for selecting those design parameters are also given so that the size of the residual set can be reduced.}
Note that the communication graphs of MASs considered in  this work is {assumed to be} undirected, which is not often {the} case in practical engineering applications. 
One possible future research topic is to investigate the fixed-time cooperative control of MASs under DoS attacks over {directed graphs.}

	\appendix	
\begin{appendices}
\section*{Proof of Theorem \ref{lemmat}}
The proof is divided into two parts. The first part is to show that the state $\epsilon_{in}$ of the observer globally converges to the state $v$ of the exosystem in a fixed time, {while the second part is to show the state $\hat v_i$ of the observer globally converges to the state $\epsilon_{in}$ in a fixed time.
{Utilizing} the obtained results in these two parts, one {can conclude}} that the state $\hat v_i$ of the observer globally converges to the state $v$ of the exosystem in a fixed time.

{\textit{Part I}}: {Show the global convergence of $\epsilon_{in}$ to $v$ in a fixed time.}

Define the estimation error $\tilde \epsilon_{in}=\epsilon_{in}-v$. By (\ref{exo}) and (\ref{obb1}), the dynamics of $\tilde \epsilon_{in}$ can be obtained as follows,
\begin{small}\begin{align}\label{obe1}
	\dot {\tilde \epsilon}_{in}=&S\tilde \epsilon_{in}-\mu_1(\sum_{j=1}^{N}a_{ij}^{\sigma(t)}(\tilde \epsilon_{in}-\tilde \epsilon_{jn})+a_{i0}^{\sigma(t)}\tilde \epsilon_{in})\nonumber\\
	&-\mu_2(\sum_{j=1}^{N}a_{ij}^{\sigma(t)}(\tilde \epsilon_{in}-\tilde \epsilon_{jn})^{\frac{a}{b}}+a_{i0}^{\sigma(t)}\tilde \epsilon_{in}^{\frac{a}{b}})\nonumber\\
	&-\mu_3(\sum_{j=1}^{N}\!a_{ij}^{\sigma(t)}(\tilde \epsilon_{in}-\tilde \epsilon_{jn})^{\frac{2b-a}{b}}+a_{i0}^{\sigma(t)}\tilde \epsilon_{in}^{\frac{2b-a}{b}}).
\end{align}\end{small}Next, the global finite-time stability of the equilibrium {point} $\tilde \epsilon_{in}\!=\!0$ of (\ref{obe1}) will be proved in two {stages}: its global finite-time convergence is shown in {\textit{Stage 1}}, and its global asymptotic stability is proved in  {\textit{Stage 2}}.

{\textit{Stage 1:}} Choose the Lyapunov function candidate $U(t)=\frac{1}{2}\sum_{i=1}^{N}\tilde \epsilon_{in}^T\tilde \epsilon_{in}$. One can see that $U(t)$ is continuous over $[0,\infty)$.
The time derivative of $U(t)$ along (\ref{obe1}) is {given as follows,}
\begin{small}\begin{align}\label{qq2}
	\dot U(t)\!=&\frac{1}{2}\!\sum_{i=1}^{N}\!\tilde \epsilon_{in}^T(S\!+\!S^T\!)\tilde \epsilon_{in}\!-\!\mu_1\!\!\sum_{i=1}^{N}\!\tilde \epsilon_{in}^T\!(\sum_{j=1}^{N}\!a_{ij}^{\sigma(t)}\!(\tilde \epsilon_{in}\!-\!\tilde \epsilon_{jn})\!\!+\!a_{i0}^{\sigma(t)}\!\tilde \epsilon_{in})\nonumber\\
	&-\mu_2\sum_{i=1}^{N}\!\tilde \epsilon_{in}^T(\sum_{j=1}^{N}a_{ij}^{\sigma(t)}(\tilde \epsilon_{in}-\tilde \epsilon_{jn})^{\frac{a}{b}}+a_{i0}^{\sigma(t)}\tilde \epsilon_{in}^{\frac{a}{b}})\nonumber\\
	&\!-\!\mu_3\!\sum_{i=1}^{N}\!\tilde \epsilon_{in}^T(\sum_{j=1}^{N}\!a_{ij}^{\sigma(t)}(\tilde \epsilon_{in}-\tilde \epsilon_{jn})^{\frac{2b-a}{b}}+a_{i0}^{\sigma(t)}\tilde \epsilon_{in}^{\frac{2b-a}{b}}).
\end{align}\end{small}

Denote the elements $l_{ij}^{\sigma(t)}$, $\bar l_{ij}^{\sigma(t)}$, $\tilde l_{ij}^{\sigma(t)}$ of {the} matrices ${\mathcal L}^{\sigma(t)}$, ${\mathcal {\bar L}}^{\sigma(t)}$, ${\mathcal {\tilde L}}^{\sigma(t)}\!\in\! \mathbb{R}^{N \!\times \!N}$ as follows: $l_{ii}^{\sigma(t)}\!=\!\sum_{j=1}^N  \!a_{ij}^{\sigma(t)}$ and $l_{ij}^{\sigma(t)}\!=\!$ $- a_{ij}^{\sigma(t)}$ for $i\neq j$, $\bar l_{ii}^{\sigma(t)}=\sum_{j=1}^N \! (a_{ij}^{\sigma(t)})^{\frac{2b}{a+b}}$ and 
$\bar l_{ij}^{\sigma(t)}\!=\!- (a_{ij}^{\sigma(t)})^{\frac{2b}{a+b}}$ for $i\neq j$, and $\tilde l_{ii}^{\sigma(t)}=\sum_{j=1}^N (a_{ij}^{\sigma(t)})^{\frac{2b}{3b-a}}$ and $\tilde l_{ij}^{\sigma(t)}=$ 
$-(a_{ij}^{\sigma(t)})^{\frac{2b}{3b-a}}$ for $i\neq j$. 
Besides, let $\mathcal H^{\sigma(t)}\! = \!diag(a_{10}^{\sigma(t)}, a_{20}^{\sigma(t)},\cdots\!,$
$a_{N0}^{\sigma(t)})\!+\!\mathcal L^{\sigma(t)}$, $\mathcal {\bar H}^{\sigma(t)} \!=\!  diag((a_{10}^{\sigma(t)})\!^{\frac{2b}{b+a}}, (a_{20}^{\sigma(t)})\!^{\frac{2b}{b+a}}, \cdots\!,(a_{N0}^{\sigma(t)})\!^{\frac{2b}{b+a}})$ 
 $+\mathcal {\bar L}^{\sigma(t)}$ and $\mathcal {\tilde H}^{\sigma(t)}\!=\! diag((a_{10}^{\sigma(t)})\!^{\frac{2b}{3b\!-\!a}}, (a_{20}^{\sigma(t)})\!^{\frac{2b}{3b\!-\!a}},\cdots\!, (a_{N0}^{\sigma(t)})\!^{\frac{2b}{3b\!-\!a}})$
 $+\mathcal {\tilde L}^{\sigma(t)}$. 
Then, by letting $\tilde \epsilon_{n}=[\tilde \epsilon_{1n}^T, \tilde \epsilon_{2n}^T, \cdots,  \tilde \epsilon_{Nn}^T]^T$ {and noting that $a_{ij}^{\sigma(t)}\!=\!a_{ji}^{\sigma(t)}$, it follows from Lemmas \ref{le1} and \ref{le2} that}
\begin{small}\begin{align}\label{qn2}
	\dot U(t)\leq
	&\frac{1}{2}\tilde \epsilon_n^T(I_N\otimes (S+S^T))\tilde \epsilon_n-\mu_1\tilde \epsilon^T_n(\mathcal H^{\sigma(t)}\otimes I_q)\tilde \epsilon_n\nonumber\\
	&-\frac{\mu_2}{2}(2\tilde \epsilon^T_n(\mathcal {\bar H}^{\sigma(t)}\otimes I_q)\tilde \epsilon_n)^{\frac{a+b}{2b}}\nonumber\\
	&-\frac{\mu_3}{2}\!(2qN^2)\!^{\frac{a-b}{2b}}\!(2\tilde \epsilon^T_n(\mathcal {\tilde H}^{\sigma(t)}\!\otimes \!I_q)\tilde \epsilon_n)\!^{\frac{3b-a}{2b}}.
\end{align}\end{small}

{Taking into consideration DoS attacks with} graph $\mathcal{\bar G}^{\sigma(t)}$, the following two cases are considered to show {
that $\epsilon_{in}$ globally converges to $v$ in a finite time.}

{\textit{Case 1}}: $t\in[t_{k-1}, t_k^a)$.

Given Assumption \ref{ass1}, it follows from Lemma \ref{lem1} that {$\mathcal H^{\sigma(t)},\mathcal {\bar H}^{\sigma(t)},$ $  \mathcal {\tilde H}^{\sigma(t)}\!>\!0$.} {It then follows from (\ref{qn2}) and condition 1) that}
\begin{small}\begin{align}\label{qn4}
	\dot U(t)
	&\leq-c_1U(t)-c_2(U(t))^{\frac{a+b}{2b}}-c_3(U(t))^{\frac{3b-a}{2b}}.
\end{align}\end{small}\quad \textit{Case 2:} $t\in[t_k^a,t_k)$.

{Even though} node $0$ might not be globally reachable {in graph $\mathcal {\bar G}^{\sigma(t)}$ during DoS attacks {time} intervals, from Lemma \ref{lem1}, one still has} $\mathcal H^{\sigma(t)}, \mathcal {\bar H}^{\sigma(t)}, \mathcal {\tilde H}^{\sigma(t)}\geq 0$.
Then, {it follows from (\ref{qn2}) that}
\begin{small}\begin{align}\label{qn5}
{\dot U(t)\leq c_4U(t).}
\end{align}\end{small}\quad Then, to show the convergence property of $U(t)$, the following two {scenarios} are considered.

{\textit{{Scenario} 1}: $U(t_0)>1$.}

In this {scenario}, the convergence property of $U(t)$ will be proved in the following two steps. In \textit{Step 1}, it will be proved that there exists $\bar t_o$ such that $U(t)\leq 1$, for $t\geq \bar t_o$. Then, together with the result obtained in \textit{Step 1}, it will be further shown in \textit{Step 2} that there exists $\tilde t_o$ such that $U(t)=0$, for $t\geq \tilde t_o$.

\textit{Step 1}: 
It follows from (\ref{qn4}) and (\ref{qn5}) that
\begin{small}\begin{subnumcases}{\label{eqn1.1}\dot U(t)\!\leq\!}
	\!-c_1U(t)\!-c_3(U(t))^{\frac{3b-a}{2b}},t\in [t_{k-1}, t^a_k),\label{eqsystem1.1}\\
	c_4U(t),t\in [t_k^a,t_{k}).\label{eqsystem2.1}
\end{subnumcases}\end{small}For $t\in [t_{k-1}, t^a_k)$, by (\ref{eqsystem1.1}), one has
\begin{small}\begin{align}\label{dv1}
	(U\!(t))\!^{\frac{a-b}{2b}}\!\geq\! ((U\!(t_{k-1}))\!^{\frac{a-b}{2b}}\!+\!\frac{c_3}{c_1}\!)e^{-\frac{c_1(a-b)}{2b}\!(t-t_{k-1})}\!-\!\frac{c_3}{c_1}.
\end{align}\end{small}For $[t_k^a,t_{k})$, by (\ref{eqsystem2.1}), one has
\begin{small}\begin{align}\label{dv2}
	(U(t))^{\frac{a-b}{2b}}\geq (U(t_k^a))^{\frac{a-b}{2b}}e^{\frac{c_4(a-b)}{2b}(t-t_k^a)}.
\end{align}\end{small}

Together with (\ref{dv1}) and (\ref{dv2}), one has\\
i) for $t\in [t_{k-1}, t^a_k)$,
\begin{small}\begin{align}\label{dv3}
	(U(t))\!^{\frac{a-b}{2b}}\!\geq&
	((U(t_{k-1}))^{\frac{a-b}{2b}}+\frac{c_3}{c_1})e^{-\frac{c_1(a-b)}{2b}(t-t_{k-1})}-\frac{c_3}{c_1}\nonumber\\
	\geq &
	((U(t_{k-1}^a))^{\frac{a-b}{2b}}e^{\frac{c_4(a-b)}{2b}(t_{k-1}-t_{k-1}^a)}+\frac{c_3}{c_1})\nonumber\\
	&\times e^{-\frac{c_1(a-b)}{2b}(t-t_{k-1})}-\frac{c_3}{c_1}\nonumber\\
	\geq& \cdots\nonumber\\
			\!\geq&((U(t_0))^{\frac{a-b}{2b}}+\frac{c_3}{c_1})e^{\frac{c_4(a\!-\!b)}{2b}\!(\!\sum_{l=1}^{k-1}\!(t_l\!-\!t_l^a))}\nonumber\\
		&\!\times\! e^{-\!\frac{c_1(a-b)}{2b}\!(\sum_{l=1}^{k-1}\!(t_l^a\!-\!t_{l\!-\!1})\!+\!(t\!-\!t_{k\!-\!1}))}	-\frac{c_3}{c_1};
\end{align}\end{small}
ii) for $[t_k^a,t_{k})$, 
\begin{small}\begin{align}\label{dv4}
	(U(t))^{\frac{a-b}{2b}}
	\geq &
	(U(t_k^a))^{\frac{a-b}{2b}}e^{\frac{c_4(a-b)}{2b}(t-t_k^a)}\nonumber\\
	\geq&
	((U(t_{k-1}))^{\frac{a-b}{2b}}\!\!+\!\frac{c_3}{c_1})e^{\frac{c_4(a-b)}{2b}(t-t_k^a)-\frac{c_1(a-b)}{2b}(t_k^a-t_{k-1})}\nonumber\\
	&-\frac{c_3}{c_1}e^{\frac{c_4(a-b)}{2b}(t-t_k^a)}\nonumber\\
	\geq& \cdots\nonumber\\%
			\geq&((U(t_0))^{\frac{a-b}{2b}}+\frac{c_3}{c_1})e^{\frac{c_4(a-b)}{2b}(\sum_{l=1}^{k\!-\!1}(t_{l}\!-\!t_{l}^a)+(t-t_k^a))\!}\nonumber\\
			&\times\! e^{-\!\frac{c_1(a-b)}{2b}(\sum_{l=1}^{k}(t_l^a\!-\!t_{l\!-\!1}))}
			-\frac{c_3}{c_1}e^{\frac{c_4(a-b)}{2b}(t-t_k^a)}.
\end{align}\end{small}{Under Assumption \ref{assdos}, it follows from (\ref{dv3}), (\ref{dv4}), $(U(t_0))\!^{\frac{a\!-\!b}{2b}}\!\geq\!0$, and $a<b$ that}
\begin{small}\begin{align}
	(U(t))^{\frac{a-b}{2b}}\!&\geq\! \frac{c_3}{c_1}(e^{\frac{(b-a)(c_1(p_d-1)-c_4)}{2bp_d}t-\frac{(c_1+c_4)(b-a)}{2b}\nu_d}\!-\!1).
\end{align}\end{small}Denote \begin{small}\begin{align}
	\label{g1}g_1(t)\!=\!\frac{c_3}{c_1}(e^{\frac{(b-a)(c_1(p_d-1)-c_4)}{2bp_d}t-\frac{(c_1+c_4)(b-a)}{2b}\nu_d}\!-\!1)\!-\!1.
\end{align}\end{small}{Then, together with condition 2), 
one has $g_1(0)<0$ and $\dot g_1(t)>0$, which imply} that there exits $\bar t_o$ such that $g_1(t)=0$. Then, one further has $(U(t))^{\frac{a-b}{2b}}\geq1$, $\forall t\geq \bar t_o$, where $\bar t_o$ is calculated by (\ref{to1}).
One thus can conclude that $U(t)\leq 1$, $\forall t\geq \bar t_o$. 

\textit{Step 2:} It follows from (\ref{qn4}) and (\ref{qn5}) that
\begin{small}\begin{subnumcases}{\label{eqn1.2}\dot U(t)\!\leq\!}
	-c_2(U(t))^{\frac{a+b}{2b}},t\in [t_{k-1}, t^a_k),\label{eqsystem1.2}\\
	c_4U(t),t\in [t_k^a,t_{k}).\label{eqsystem2.2}
\end{subnumcases}\end{small}For $t\in [t_{k-1}, t^a_k)$, by (\ref{eqsystem1.2}), one has
\begin{small}\begin{align}\label{dv5}
	(U(t))^{\frac{b-a}{2b}}\!\leq (U(t_{k-1}))^{\frac{b-a}{2b}}-\frac{c_2(b-a)}{2b}(t-t_{k-1}).
\end{align}\end{small}For $t\in [t_k^a,t_{k})$, by (\ref{eqsystem2.2}), one has
\begin{small}\begin{align}\label{dv6}
	(U(t))^{\frac{b-a}{2b}}\leq (U(t_k^a))^{\frac{b-a}{2b}}e^{\frac{c_4(b-a)}{2b}(t-t_k^a)}.
\end{align}\end{small}For the convenience of description, let $[\bar t_{k-1}, \bar t^a_k)$ and $[\bar t^a_k , \bar t_k)$ be the {$k$th} time intervals for graph $\mathcal {\bar G}^{\sigma(t)}$ without DoS attacks and under DoS attacks, respectively, where $\bar t_0=\bar t_o$.


It follows from (\ref{dv5}) and (\ref{dv6}) that 

i) for $[\bar t_{k-1}, \bar t^a_k)$,
\begin{small}\begin{align}\label{dv7}
	(U(t))^{\frac{b-a}{2b}}
	\leq& (U(\bar t_{k-1}))^{\frac{b-a}{2b}}-\frac{c_2(b-a)}{2b}(t-\bar t_{k-1})\nonumber\\
	\leq &( U(\bar t_{k-1}^a))^{\frac{b\!-\!a}{2b}}\!e^{\frac{c_4(b\!-\!a)}{2b}(\bar t_{k\!-\!1}-\bar t_{k\!-\!1}^a)}\!-\!\!\frac{c_2(b\!-\!a)}{2b}\!(t\!-\!\bar t_{k\!-\!1})\nonumber\\
	\leq& \cdots\nonumber\\
	\leq& (U(\bar t_o))^{\frac{b-a}{2b}}e^{\frac{c_4(b-a)}{2b}(\sum_{l=1}^{k-1}(\bar t_l-\bar t_l^a))}\nonumber\\
	&\!-\!\frac{c_2(b-a)}{2b}(\sum_{l=1}^{k-1}\!(\bar t_{l}-\bar t_{l}^a)+(t-\bar t_{k-1}));
\end{align}\end{small}
ii) for $[\bar t_{k}^a, \bar t_k)$,
\begin{small}\begin{align}\label{dv8}
	(U(t))^{\frac{b-a}{2b}}\leq&(U(\bar t_k^a))^{\frac{b-a}{2b}}e^{\frac{c_4(b-a)}{2b}(t-\bar t_{k}^a)}\nonumber\\
	\leq&(U(\bar t_{k-1}))^{\frac{b-a}{2b}}e^{\frac{c_4(b-a)}{2b}(t-\bar t_{k}^a)}\nonumber\\
	&-\frac{{c_2}(b-a)}{2b}e^{\frac{c_4(b-a)}{2b}(t-\bar t_{k}^a)}(\bar t_k^a-\bar t_{k-1})\nonumber\\
	\leq &\cdots\nonumber\\
	\leq&(U(\bar t_{o}))^{\frac{b-a}{2b}}e^{\frac{c_4(b-a)}{2b}((t-\bar t_{k}^a)+\sum_{l=1}^{k-1}(\bar t_{l}-\bar t_{l}^a))}\nonumber\\
	&-\frac{c_2(b-a)}{2b}\sum_{l=1}^{k}(\bar t_{l}^a-\bar t_{l-1}).
\end{align}\end{small}Note that $U(\bar t_o)\leq 1$ from {\textit{Step 1}}. {Under Assumption \ref{assdos}, 
it follows from (\ref{dv7}) and (\ref{dv8}) that
\begin{small}\begin{equation}\label{dotfixed}
	(U(t))^{\frac{b-a}{2b}}\!\leq 
	e^{\frac{c_4\!(b-a)}{2b}\!(\!\frac{t-\bar t_o}{p_d}+\nu_d\!)}
\!-\!\frac{c_2(b\!-\!a)}{2b}(t\!-\!\bar t_o\!-\!\frac{t\!-\!\bar t_o}{p_d}\!-\!\nu_d).
\end{equation}\end{small}Denote
	$g_2(t)=e^{\frac{c_4(b-a)}{2b}(\frac{t-\bar t_o}{p_d}+\nu_d)}-\frac{c_2(b-a)}{2b}(t-\bar t_o-\frac{t-\bar t_o}{p_d}-\nu_d).$}
One then has $\dot g_2(t)=\frac{c_4(b-a)}{2bp_d}e^{\frac{c_4(b-a)}{2b}(\frac{t-\bar t_o}{p_d}+\nu_d)}-\frac{c_2(b-a)}{2b}(1-\frac{1}{p_d})$ and $\ddot g_2(t)=(\frac{c_4(b-a)}{2bp_d})^2e^{\frac{c_4(b-a)}{2b}(\frac{t-\bar t_o}{p_d}+\nu_d)}$.
It follows from conditions 3) and 4) that $g_2(\tilde t_o)$ is the minimal value of $g_2(t)$ and $g(\tilde t_o)\leq 0$, where $\tilde t_o$ is calculated by (\ref{ton}).
Then, it follows from (\ref{dotfixed}) that $(U(\tilde t_o))^{\frac{b-a}{2b}}\leq 0$. Since $U(t)\geq 0$, $\forall t\geq 0$, it can be concluded that $U({\tilde t_o})=0$.

{\textit{{Scenario} 2}: $U(t_0)\leq1$}.

In this {scenario}, from the analysis in \textit{Step 2} of \textit{{Scenario} 1}, one can infer that $U(\hat t_o)=0$, where $\hat t_o=\tilde t_o-\bar t_o$.

Therefore, combining the results obtained in the above two {scenarios}, one has for any $\tilde \epsilon_{in}(0)\in {\mathbb R^q}$, $\tilde \epsilon_{in}(\tilde t_o)=0, i=1,2,\cdots,N$. The global finite-time convergence of the equilibrium {point} $\tilde \epsilon_{in}=0$ is thus proved.

{\textit{Stage 2:}} It follows from  (\ref{eqn1.1})  in {\textit{Stage 1} that}
\begin{small}\begin{align}\label{vc}
	\dot U(t)\leq c_4U(t), \forall t\geq 0.
\end{align}\end{small}Integrating both sides of inequality (\ref{vc}) over $[0,t)$ leads to
\begin{small}\begin{align}
	U(t)\leq e^{c_4t}U(0),  \forall t\geq 0.
\end{align}\end{small}For $ t\in[0, \tilde t_o)$, one further has $U(t)\leq e^{c_4\tilde t_o}U(0)$.
Since $U(t)=\frac{1}{2}\sum_{i=1}^{N}\tilde \epsilon_{in}^T\tilde \epsilon_{in}$, one has
$\|\tilde \epsilon_n\|^2\leq e^{c_4\tilde t_o}\|\tilde \epsilon_n(0)\|^2$, $\forall t\in[0, \tilde t_o)$. 
Thus, one can infer that for any $\varepsilon>0$, there is $\delta(\varepsilon)>0$ such that for every $ \|\tilde \epsilon_n(0)\|\leq \varepsilon$, 
$\|\tilde \epsilon_n\|\leq \delta$. Then, together with $\lim\limits_{t\to \tilde t_o} \tilde \epsilon_n=0$, $\forall \tilde \epsilon_n(0)\in {\mathbb R^q}$ obtained in {\textit{Stage 1}}, one can conclude that $\tilde \epsilon_{in}=0$ is globally asymptotically stable.

According to the results obtained in {\textit{Stage 1} and \textit{Stage 2}}, along with Definition \ref{stable} (i), one can conclude that the equilibrium {point} $\tilde \epsilon_{in}=0$ of (\ref{obe1}) is globally finite-time stable. 
Moreover, the settling time is bounded by $\tilde t_o$, which is independent of the initial conditions.
Therefore, by Definition \ref{stable} (ii), it can be concluded that the equilibrium {point} $\tilde \epsilon_{in}=0$ of (\ref{obe1}) is globally fixed-time stable. Since $\tilde \epsilon_{in}=\epsilon_{in}-v$, it can be concluded that $\epsilon_{in}$ globally converges to $v$  in fixed time $\tilde t_o$.

{\textit{Part II}}: Show that $\hat v_i$ globally converges to {$\epsilon_{in}$} in a fixed time through the following steps.

\textit{Step 1}. 
Let $\tilde \epsilon_{i(n\!-\!1)}\!=\!\epsilon_{i(n\!-\!1)}\!\!-\!\epsilon_{in}$. It follows from (\ref{obb4}) and (\ref{obb2}) that 
\begin{small}\begin{align}\label{obbv1}
	\dot{\tilde \epsilon}_{i(n-1)}\!=&S \tilde \epsilon_{i(n-1)}\!-\!\delta_1\tilde \epsilon_{i(n-1)}\!-\!\delta_2{\tilde \epsilon_{i(n-1)}}^{\frac{a}{b}}\!-\!\delta_3{\tilde \epsilon_{i(n-1)}}^{2-\frac{a}{b}}+\Delta_{i},
\end{align}\end{small}where $\Delta_{i}\!=\!\mu_1\!\sum_{j=0}^{N}\!a_{ij}^{\sigma(t)}(\epsilon_{in}-\epsilon_{jn})+\mu_2\sum_{j=0}^{N}a_{ij}^{\sigma(t)} (\epsilon_{in}\!-\!\epsilon_{jn})^{\frac{a}{b}}$ $+\mu_3\sum_{j=0}^{N}a_{ij}^{\sigma(t)}(\epsilon_{in}-\epsilon_{jn})^{2-\frac{a}{b}}$. 
Since $\epsilon_{in}$ globally converges to $v$  in fixed time $\tilde t_o$, $\Delta_{i}$ globally converges to zero in a fixed time $\tilde t_o$. Thus, (\ref{obbv1}) reduces to
\begin{small}\begin{align}\label{obbv2}
	\dot{\tilde \epsilon}_{i(n-1)}\!\!=&S \tilde \epsilon_{i(n-1)}\!\!-\!\delta_1\tilde \epsilon_{i(n-1)}\!\!-\!\delta_2\tilde \epsilon_{i(n-1)}^{\frac{a}{b}}\!-\!\delta_3\tilde \epsilon_{i(n-1)}^{2-\frac{a}{b}}, \forall t\!\geq\!\tilde t_o.
\end{align}\end{small}The Lyapunov function candidate in this step is chosen as $W_1(t)\!=\!\frac{1}{2}\!\sum_{i=1}^{N}\!\tilde \epsilon_{i(n-1)}^T\tilde \epsilon_{i(n-1)}$. 
{Then, it follows from condition 5) and Lemma \ref{le1} that the time derivative of $W_1(t)$ along (\ref{obbv2}) satisfies
}\begin{small}\begin{align}
	\dot W_1(t)\leq -\delta_2(W_1(t))^{\frac{a+b}{2b}}-\delta_3(Nq)^{\frac{a-b}{2b}}(W_1(t))^{\frac{3b-a}{2b}}.
\end{align}\end{small}By Lemma \ref{fixed time stability}, it can be inferred that $\epsilon_{i(n-1)}$ globally converges to $\epsilon_{in}$ in a fixed time  with its upper bound giving by $t_{ o1}=\tilde t_{o}+{\frac{2b}{(b-a)}(\frac{1}{\delta_2}+\frac{1}{\delta_3(Nq)^{\frac{a-b}{2b}}})}$.

\textit{Step $s$. $(s\!=\!2,\cdots,n-1)$}. Define $\tilde \epsilon_{i(n-s)}\!=\! \epsilon_{i(n-s)}\!-\!\epsilon_{i(n-s+1)}$. 
With the result obtained in \textit{Step $s\!-\!1$} in this part and  (\ref{ob}), one has
\begin{small}\begin{align}\label{obbvs}
	\dot{\tilde \epsilon}_{i(n\!-\!s)}\!\!=&S \tilde \epsilon_{i(n\!-\!s)}\!\!-\!\delta_1\tilde \epsilon_{i(n\!-\!s)}\!\!-\!\delta_2\tilde \epsilon_{i(n\!-\!s)}^{\frac{a}{b}}\!\!-\!\delta_3\tilde \epsilon_{i(n\!-\!s)}^{2-\frac{a}{b}}, \forall t\!\geq \!t_{o(s\!-\!1)},
\end{align}\end{small}where $t_{o(s\!-\!1)}\!=\tilde t_{o}+{\!\frac{2b(s\!-\!1)}{b-a}(\frac{1}{\delta_2}+\frac{1}{\delta_3(Nq)\!^{\frac{a\!-\!b}{2b}}})}$. 
The Lyapunov function candidate in this step is considered as $W_s(t)=\frac{1}{2}\sum_{i=1}^{N}\tilde \epsilon_{i(n-s)}^T\tilde \epsilon_{i(n-s)}$. 
Similar to \textit{Step 1} {in this part}, one has
\begin{small}\begin{align}
	\dot W_s(t)\leq -\delta_2(W_s(t))^{\frac{a+b}{2b}}-\delta_3(Nq)^{\frac{a-b}{2b}}(W_s(t))^{\frac{3b-a}{2b}}.
\end{align}\end{small}By Lemma \ref{fixed time stability}, one can infer that $\epsilon_{i(n-s)}$ globally converges to $\epsilon_{i(n-s+1)}$ in a fixed time
with its upper bound giving by 
$t_{ os}\!=\tilde t_{o}+{\!\frac{2bs}{b-a}(\frac{1}{\delta_2}+\frac{1}{\delta_3(Nq)\!^{\frac{a\!-\!b}{2b}}})}$.

\textit{Step $n$}. Define $\tilde v_i=\hat v_i-\epsilon_{i1}$. According to (\ref{ob}), one has
\begin{small}\begin{align}\label{obbvn}
	\dot{\tilde v}_i=&S \tilde v_i-\delta_1\tilde v_i-\delta_2\tilde v_i^{\frac{a}{b}}
-\delta_3\tilde v_i^{2-\frac{a}{b}}, \forall t\geq t_{o(n-1)},
\end{align}\end{small}where $t_{o(n-1)}\!=\tilde t_{o}+{\!\frac{2b(n\!-\!1)}{b-a}(\frac{1}{\delta_2}+\frac{1}{\delta_3(Nq)\!^{\frac{a\!-\!b}{2b}}})}$.
Select the Lyapunov function candidate in this step as $W_n(t)=\frac{1}{2}\sum_{i=1}^{N}\tilde v_i^T\tilde v_i$. 
Similar to \textit{Step 1} {in this part}, it can be obtained that
\begin{small}\begin{align}
	\dot W_n(t)\leq -\delta_2(W_n(t))^{\frac{a+b}{2b}}-\delta_3(Nq)^{\frac{a-b}{2b}}(W_n(t))^{\frac{3b-a}{2b}}.
\end{align}\end{small}By Lemma \ref{fixed time stability}, {one can infer that $\hat v_i$ globally converges to $\epsilon_{i1}$ in a fixed time $t_{o}$, where $t_o$ is defined by (\ref{to}). It thus can be concluded that $\hat v_i$ globally converges to $\epsilon_{in}$ in a fixed time $t_{o}$.}

Therefore, with the results established in {\textit{Part I}} and {\textit{Part II}},  one can infer that the state $\hat v_i$ of the observer (\ref{ob}) globally converges to the state $v$ of the exosystem in a fixed time $t_{o}$.
The proof is thus completed. 
\end{appendices}

%

%
%
%


\ifCLASSOPTIONcaptionsoff
  \newpage
\fi

\renewcommand\refname{References}
\footnotesize
\scriptsize
\bibliographystyle{IEEEtran}
\bibliography{IEEEabrv,referencesuncertainleader}

\end{document}